\documentclass{article}

\usepackage{spconf}            % Pour le style
\usepackage{slashbox}
\usepackage{amsmath}
\usepackage{graphicx}
\usepackage{epsfig}
\usepackage{appendix}

\newcommand{\A}{{\bf A}}

\newcommand{\RR}{{{\rm I}\mkern-4mu{\rm R}}}

\newcommand{\CC}{{\mathchoice {\setbox0=\hbox{$\displaystyle\rm C$}\hbox{\hbox
to0pt{\kern0.4\wd0\vrule height0.9\ht0\hss}\box0}}
{\setbox0=\hbox{$\textstyle\rm C$}\hbox{\hbox
to0pt{\kern0.4\wd0\vrule height0.9\ht0\hss}\box0}}
{\setbox0=\hbox{$\scriptstyle\rm C$}\hbox{\hbox
to0pt{\kern0.4\wd0\vrule height0.9\ht0\hss}\box0}}
{\setbox0=\hbox{$\scriptscriptstyle\rm C$}\hbox{\hbox
to0pt{\kern0.4\wd0\vrule height0.9\ht0\hss}\box0}}}}

\title{A new algorithm for complex non orthogonal joint diagonalization based on Shear and Givens rotations}
\name{Mesloub Ammar$^{\star}$ \qquad Karim Abed-Meraim$^{\star\star}$ \qquad Adel Belouchrani$^{\star\star\star}$}
\address{$^{\star}$Ecole Militaire Polytechnique, BP 17 Bordj El Bahri, Algiers, Algeria\\
	  $^{\star\star}$Polytech' Orl\'eans/PRISME Lab., Univ. Orl\'eans  12 rue de Blois, 45067 Orl\'eans, France\\
    $^{\star\star\star}$Ecole Nationale Polytechnique, 10 Avenue Hassen Badi, 16200 Algiers, Algeria\\
    mesloub.a@gmail.com, karim.abed-meraim@univ-orleans.fr, adel.belouchrani@enp.edu.dz }

\begin{document}
%\begin{onecolumn}
\ninept
\maketitle
\begin{abstract}

This paper introduces a new algorithm to approximate non orthogonal joint diagonalization (NOJD) of a set of complex matrices. This algorithm is based on the Frobenius norm formulation of the JD problem and takes advantage from combining Givens and Shear rotations to attempt the approximate joint diagonalization (JD). It represents a non trivial generalization of the JDi (Joint Diagonalization) algorithm (Souloumiac 2009) to the complex case. The JDi is first slightly modified then generalized to the CJDi (i.e. Complex JDi) using complex to real matrix transformation.
Also, since several methods exist already in the literature, we propose herein a brief overview of existing NOJD algorithms then we provide an extensive comparative study to illustrate the effectiveness and stability of the CJDi w.r.t. various system parameters and application contexts. %Firstly, we have transformed complex matrices to the real onces in order to use an existing algorithm for real JD. Secondely, by observing real matrices structure, we have have developed our new algorithm (referred as CJDi Complex Joint Diagonalization) by considering the complex matrices and improved performances in terms of computational cost, convergence rate and JD quality. Numerical simulations illustrate the efficiency and stability of proposed algorithms and compared with respect to some existing  non orthogonal JD methods (ACDC, FAJD, QDiag, UWEDGE, JUST, CVFFDiag, ALS and LUCJD). 
	
\end{abstract}

%------------------------------------------------------------
%------------------------------------------------------------
\section{Introduction} \label{intro}
%------------------------------------------------------------
%------------------------------------------------------------

Joint diagonalization problem and its related algorithms are found in various applications, especially in blind source separation (BSS) and independent component analysis (ICA). In such problems, it is desired to diagonalize simultaneously a set of square matrices. These matrices can be covariance matrices estimated on different time windows \cite{Pham}, intercorrelation matrices with time shifts \cite{Beloumera}, fourth or higher order cumulant slice matrices \cite{Soulcard}\cite{Soulcard2} or spatial time-frequency matrices \cite{Bous}\cite{fada}. 
\\Mathematically, the joint diagonalization problem can be stated as follows: Given a set of $K$ square matrices $\left\{\textbf{M}_k  \in \CC^{N \times N},\right.$ $\left. ~~~ k = 1,...,K \right\}$, find a matrix $\textbf{V}$ such that the transformed matrices $\textbf{D}_k = \textbf{V} \textbf{M}_k \textbf{V}^H$ are as diagonal as possible.
\\In the context of BSS, $\left\{\textbf{M}_k\right\}$  are complex matrices sharing the same structure defined by $\textbf{M}_k = \textbf{A} \textbf{D}_k \textbf{A}^H$ where 
$\left\{\textbf{D}_k\right\}$ are diagonal matrices and $\textbf{A}$ is an unknown mixing matrix. The problem consists of finding the diagonalizing matrix $\textbf{V}$ that left inverts $\textbf{A}$ and transforms $\textbf{V} \textbf{M}_k \textbf{V}^H$ into diagonal matrices.

Various algorithms have been developed to solve the JD problem. These algorithms can be classified in two classes, orthogonal joint diagonalization (OJD) and non orthogonal JD (NOJD). The first class imposes $\textbf{V}$ to be orthogonal by transforming the JD problem into an OJD problem using the whitening step \cite{Beloumera}\cite{LSOJD}. This step can introduce errors which might reduce the diagonalization performance \cite{Card}\cite{soulou3}. The most popular algorithm for OJD is JADE \cite{JADE} which is a Jacobi-like algorithm based on Givens rotations.\\
The NOJD class treats the problem without any whitening step. Among the first NOJD algorithms, one can cite the Iterative Decorrelation Algorithm (IDA) developed for complex valued NOJD in \cite{IDA}\cite{IDA2} and the  AC-DC (Alternating Columns-Diagonal Centers) given in \cite{ACDC}. The latter suffers from slow linear convergence performance. Many other algorithms have been developed by considering specific criteria or constraints in order to avoid trivial and degenerate solutions \cite{UWEDGE}\cite{COMP}. These algorithms can be listed as follow: QDiag \cite{qdiag} (Quadratic Diagonalization algorithm) developed by Vollgraf and Obermayer where the JD criterion is rearranged as a quadratic cost function; FAJD \cite{FAJD} (Fast Approximative Joint Diagonalization)  developed by Li and Zhang where the diagonalizing matrix is estimated column by column; UWEDGE \cite{UWEDGE} (UnWeighted Exhaustive joint Diagonalization with Gauss itErations) developed by Tichavsky and Yeredor where numerical optimization is used to get the JD solution; JUST \cite{JUST} (Joint Unitary and Shear Transformations) developed by Iferroudjene, Abed-Meraim and Belouchrani where the algebraic joint diagonalization is considered; CVFFDiag \cite{CVFFDiag}\cite{FFDiag} (Complex Valued Fast Frobenius Diagonalization) developed by Xu, Feng and Zheng where first order of Taylor expansion is used to minimize the JD criterion; ALS \cite{als} (Alternating Least Squares) developed by Trainini and Moreau where the mixing and diagonal matrices are estimated alternatively by using least squares criterion and LUCJD \cite{LUCJD}\cite{Afsari} (LU decomposition for Complex Joint Diagonalization) developed by Wang, Gong and Lin where the diagonalizing matrix is estimated by LU decomposition.

In this paper, we generalize the JDi algorithm  developed by Souloumiac in \cite{JDi} for real joint diagonalization by using Shear and Givens rotations in the complex case\footnote{A first attempt to generalize the JDi has been given in \cite{feng}. Unfortunately, the latter algorithm  has been found to diverge in most simulation contexts considered in section \ref{simul}, and hence, it has been omitted in our comparative study.}. We transform the considered complex matrices to real symmetric ones allowing us to apply the JDi algorithm. At the convergence, the diagonalizing (mixing) matrix is retrieved from the real diagonalizing one by taking into account the particular structure of the latter (see subsection \ref{CDMR} for more details). The main drawback of this algorithm's version is that it does not take into consideration the particular structure of the real valued diagonalizing matrix along the iterations which results in a slight performance loss.  To avoid this drawback, we propose an improved version which uses explicitly the complex matrices using special structure of Shear and Givens rotations to increase both the convergence rate and the estimation accuracy, while reducing the overall computational cost. 
Another contribution of this paper is a comparative study of different non orthogonal joint diagonalization algorithms with respect to their robustness in severe JD conditions: i.e. large dimensional matrices, noisy matrices, ill conditioned matrices and large valued MOU (modulus of uniqueness \cite{JDi}\cite{Afs}\cite{shearnote}). 

The rest of the paper is organized as follows. In section \ref{prob}, the problem formulation, mathematical notations and paper's main objectives are stated.  Section \ref{NOJD} introduces a brief overview of major NOJD algorithms and existing JD criteria. Section \ref{G_JDi} presents the basic generalization of JDi algorithm to the complex case while section \ref{cjdi} is dedicated to the proposed method's developments. In particular, we present in this section  a complex implementation of our method with a computational cost comparison with existing NOJD algorithms. Simulation based performance assessment for exact and approximate joint diagonalizable matrices are provided in section \ref{simul}. Section \ref{conclusion} is dedicated to the concluding remarks.  

%Section \ref{cjdi}, presents a brief overview of the main existing NOJD methods.

%------------------------------------------------------------
%------------------------------------------------------------
\section{Problem formulation} \label{prob}
%------------------------------------------------------------
%------------------------------------------------------------
%This section containes mathematic problem difinition and algorithms' steps and developpement.
Consider a set of $K$ square matrices,  $\textbf{M}_k  \in \CC^{N \times N} ~~~ k = 1,...,K $ sharing the following decomposition:   
\begin{equation}
\begin{array}{cc} \textbf{M}_k=\textbf{A} \textbf{D}_k \textbf{A}^H & ,k=1,\cdots,K \end{array}
\label{eq1}
\end{equation}
where $\textbf{D}_k$ are diagonal matrices and $\textbf{A}$ is the unknown $N \times N$ square complex non-defective matrix known as a mixing matrix. $\textbf{A}^H$ denotes the transpose conjugate of $\textbf{A}$. \\
The problem of joint diagonalization consists of estimating matrices $\textbf{A}$ and $\textbf{D}_k$, $k=1,\cdots,K$ given the observed matrices $\textbf{M}_k$, $k=1,\cdots,K$.  Equivalently, the JD problem consists of finding the transformation $\textbf{V}$ such that matrices $\textbf{V} \textbf{M}_k \textbf{V}^H$ are diagonal. 
Note that JD decomposition given in equation (\ref{eq1}) is not unique. Indeed,  if $\left\{ \textbf{A}, \textbf{D}_1, \cdots, \textbf{D}_K\right\}$ is a solution, then for any permutation matrix $ \textbf{P} \in \RR^{N \times N}$  and invertible diagonal matrix $\textbf{D}\in \CC^{N \times N}$,   $\left\{ \textbf{A} \textbf{D} \textbf{P}, \textbf{P}^T \textbf{D}^{-1} \textbf{D}_1 \textbf{D}^{-H} \textbf{P}\right.$, $\left.\cdots, \textbf{P}^T \textbf{D}^{-1}\textbf{D}_K \textbf{D}^{-H}\textbf{P}\right\}$ is also a solution.  Fortunately, in most practical applications these indeterminacies are inherent and do not affect the final result of the considered problem.
In practice, matrices $\textbf{M}_1, \cdots,\textbf{M}_K$ are given by some sample averaged statistics that are corrupted by estimation errors due to noise and finite sample size effects. Thus, they are only "approximately" jointly diagonalizable matrices and can be rewritten as:
\begin{equation}
\begin{array}{cc} \textbf{M}_k = \textbf{A} \textbf{D}_k \textbf{A}^H + \Xi_k & ,k=1,\cdots,K \end{array}
\label{eq2}
\end{equation}
where $\Xi_k$ are perturbation (noise) matrices.  

%------------------------------------------------------------
%------------------------------------------------------------
\section{Review of major NOJD algorithms} \label{NOJD}
%------------------------------------------------------------
%------------------------------------------------------------
In this section, we present a brief overview of NOJD algorithms. First, the different JD criteria are presented before giving a brief description for each considered algorithm. 

%------------------------------------------------------------
%------------------------------------------------------------
\subsection{Joint diagonalization criteria} \label{lsa}
%------------------------------------------------------------
%------------------------------------------------------------
In this subsection, we present different criteria considered for JD problem. The first one is given in \cite{ACDC} and expressed as follow:
\begin{equation}
C_{DLS}(\textbf{A},\textbf{D}_1,\textbf{D}_2,...,\textbf{D}_K)=\sum_{k=1}^{K}{w_k\left\| \textbf{M}_k - \textbf{A} \textbf{D}_k \textbf{A}^H \right\|_F^2} 
\label{cdls}
\end{equation}
where $\left\|.\right\|_F$ refers to the Frobenius norm, $w_k$ are some positive weights and $(\textbf{A},\textbf{D}_1,\textbf{D}_2,...,\textbf{D}_K)$ are the searched mixing matrix and diagonal matrices, respectively. This cost function is called in \cite{DIEM}, the Direct Least-Squares (DLS) criterion as it takes into account the mixing matrix rather than the diagonalizing one.\\
Unlike the previous JD criterion, the second one is called the Indirect Least Squares (ILS) criterion and takes into account the diagonalizing matrix $\textbf{V}$. The latter is expressed as \cite{DIEM}:

\begin{equation}
 C_{ILS}(\textbf{V})=\sum_{k=1}^{K}{\left\| \textbf{V} \textbf{M}_k \textbf{V}^H - \mbox{diag}\left( \textbf{V} \textbf{M}_k \textbf{V}^H \right) \right\|_F^2} 
\label{cils}
\end{equation}

This criterion is widely used in numerous algorithms \cite{Beloumera}\cite{UWEDGE}\cite{Afs}. However, the minimization of (\ref{cils}) might lead to undesired solutions e.g. trivial solution $\textbf{V}=\textbf{0}$ or degenerate solutions where $\mbox{det}(\textbf{V}) = 0 $. Consequently, the algorithms based on the minimization of (\ref{cils}) introduce a constraint to avoid these undesirable solutions. In \cite{Beloumera}, the estimated mixing matrix (resp. the diagonalizing matrix) have to be unitary so that undesired solutions are avoided. In our developed algorithm, the diagonalizing matrix is estimated as a product of Givens and Shear rotations where undesired solutions are excluded implicitly. In \cite{LUCJD}, the diagonalizing matrix $\textbf{V}$ is estimated in the form of LU (or LQ) factorization where $\textbf{L}$ and $\textbf{U}$ are lower and upper triangular matrices with ones at the diagonals and $\textbf{Q}$ is a unitary matrix. These two factorizations impose a unit valued determinant for the diagonalizing matrix.  Previous factorizations (Givens , Givens and  shear, LU and LQ factorizations) represent the different constraints used to avoid undesired solutions.\\
In \cite{FAJD}, the undesired solutions are excluded  by considering the penalization term $-\beta \mbox{Log}(\left|\mbox{det}(\textbf{V})\right|)$ so that the JD criterion becomes:

\begin{equation}
C'_{ILS}(\textbf{V})= \sum_{k=1}^{K}{\left\| \textbf{V} \textbf{M}_k \textbf{V}^H - \mbox{diag}\left( \textbf{V} \textbf{M}_k \textbf{V}^H \right) \right\|_F^2} -\beta \mbox{Log}(\left| \mbox{det}(\textbf{V}) \right|)
\label{cmils}
\end{equation}

Another criterion has been introduced in \cite{UWEDGE} taking into account two matrices ($\textbf{V}$,$\textbf{A}$) which are the diagonalizing matrix and its residual mixing one, respectively. It is expressed as:   
\begin{equation}
C_{LS}(\textbf{V},\textbf{A} )= \sum_{k=1}^{K}{\left\| \textbf{V} \textbf{M}_k \textbf{V}^H - \textbf{A}\mbox{ diag}\left( \textbf{V} \textbf{M}_k \textbf{V}^H \right)\textbf{A}^H \right\|_F^2} 
\label{cls}
\end{equation}
The previous criterion $C_{LS}$ fuses the direct and indirect forms by relaxing the dependency between $\textbf{A}$ and $\textbf{V}$ and it is  known as least squares criterion.
In \cite{Pham}, another criterion is developed for positive definite target matrices as follow:   
\begin{equation}
C_{LL}(\textbf{V})= \sum_{k=1}^{K}{\mbox{Log}\frac{\mbox{det}\left(\mbox{diag}\left( \textbf{V} \textbf{M}_k \textbf{V}^H \right)\right)}{\mbox{det}(\textbf{V} \textbf{M}_k \textbf{V}^H)}} 
\label{cll}
\end{equation}
This criterion can not be applied for non positive target matrices, thus some real life applications can not be treated by minimizing (\ref{cll}).
In \cite{Afsari}, a scale invariant criterion in $\textbf{V}$ is introduced as: 
\begin{equation}
C_{ILS2}(\textbf{V})= \sum_{k=1}^{K}{\left\|  \textbf{M}_k  - \textbf{V}^{-1}\mbox{ diag}\left( \textbf{V} \textbf{M}_k \textbf{V}^H \right)\textbf{V}^{-H} \right\|_F^2} 
\label{cils2}
\end{equation}
Note that $C_{ILS2}(\textbf{DV})=C_{ILS2}(\textbf{V})$ for any diagonal matrix $\textbf{D}$. This criterion is used only for real JD.

%-----------------------------------------
\subsection{NOJD Algorithms}
%-----------------------------------------

We describe herein the basic principles of each of  major NOJD algorithms considered in our comparative study given in section \ref{simul}. \\[0.5cm]
%
%\begin{description}
%------------------------------------------------------------
\underline{\bf ACDC \cite{ACDC}}:
%------------------------------------------------------------
This algorithm is developed by Yeredor in 2002. It proceeds by minimizing $C_{DLS}$ criterion given in (\ref{cdls}) by alternating two steps: The first one is the AC (Alternating Columns) step and the second one is the DC (Diagonal Centers) step.  For AC step, only one column in the mixing matrix is updated by minimizing the cited criterion while the other parameters are kept fixed. For the DC step, the diagonal matrices entries are estimated by keeping the mixing  matrix fixed.
Note that, the DC phase is followed by several AC phases in order to guarantee the algorithm's convergence.\\[0.5cm]
% 
%------------------------------------------------------------
\underline{\bf FAJD \cite{FAJD}}: 
%------------------------------------------------------------
This algorithm is developed by Li and Zhang in 2007. It estimates the diagonalizing matrix by minimizing the modified indirect least squares $C'_{ILS}$ criterion given in (\ref{cmils}). At each iteration, the algorithm updates one column of the diagonalizing matrix while keeping the others fixed. This process is repeated until reaching the convergence state. Note that  the value assigned to $\beta$ in \cite{FAJD} is one.\\[0.5cm]
%
%------------------------------------------------------------
\underline{\bf QDiag \cite{qdiag}}: 
%------------------------------------------------------------
This algorithm is developed by Vollgraf and Obermayer in 2006.
It minimizes the indirect least squares criterion given in (\ref{cils}). At each iteration, the algorithm updates one column of the diagonalizing matrix while the others are kept fixed. This step is repeated until reaching the convergence state. Note that there is no update step for target matrices and the condition to avoid undesired solutions is implicitly included by normalizing the diagonalizing matrix columns.\\[0.5cm]
%
%------------------------------------------------------------
\underline{\bf UWEDGE \cite{UWEDGE}}: 
%------------------------------------------------------------
%
This algorithm is developed by Tichavsky and Yeredor in 2008. It minimizes the $C_{LS}$ criterion given in (\ref{cls}) and computes in alternative way the residual mixing and diagonalizing matrices. At first, the diagonalizing matrix $\textbf{V}$ is initialized as $\textbf{M}_1^{-\frac{1}{2}}$ \footnote{This initial value of $\textbf{V}$ is known as the whitening matrix in BSS context and assumes that $\textbf{M}_1$ is positive definite. Otherwise, other initializations can be considered.}. This value is introduced in the considered criterion to find the mixing matrix $\textbf{A}$. The minimization w.r.t. $\textbf{A}$ is achieved by using numerical Gauss iterations. Once an estimate $\hat{\textbf{A}}$ of the mixing matrix is obtained, the diagonalizing matrix is updated as $\textbf{V}^{(i)}=\hat{\textbf{A}}^{-1}\textbf{V}^{(i-1)}$ ($i$ is the iteration index). The previous process is repeated  until the convergence is reached. \\[0.5cm]
%
%------------------------------------------------------------
\underline{\bf JUST \cite{JUST}}: 
%------------------------------------------------------------
This algorithm is developed by Iferroudjene, Abed-Meraim  and in Belouchrani 2009. It is applied to target matrices sharing the algebraic joint diagonalization structure $\textbf{M}'_k=\textbf{A}\textbf{D}'_k\textbf{A}^{-1}$. Hence in our context, given the target matrices sharing the decomposition described in (\ref{eq1}). The latter are transformed to another set of new target matrices sharing the algebraic joint diagonalization structure by  right multiplying them by the inverted first target matrix. Once the new set of target matrices is obtained, JUST algorithm estimates the diagonalizing matrix by successive Shear and Givens rotations minimizing $C_{ILS}$ criterion\footnote{The considered $C_{ILS}$ in JUST can be expressed as in (\ref{cils}) by replacing $\textbf{V}^H$ by $\textbf{V}^{-1}$.}. \\[0.5cm]
%
%------------------------------------------------------------
\underline{\bf CV FFDiag \cite{CVFFDiag}}: 
%------------------------------------------------------------
This algorithm's idea is given in \cite{IDA} and it is formulated as an algorithm for real NOJD in \cite{FFDiag}. Xu, Feng and  Zheng generalized the latter to the complex NOJD in  \cite{CVFFDiag}. It minimizes $C_{ILS}$ criterion and estimates the diagonalizing matrix $\textbf{V}$ in an iterative scheme using  the following form $\textbf{V}^{(i)}=\left(\textbf{I}+\textbf{W}^{(i)}\right)\textbf{V}^{(i-1)}$ where  $\textbf{W}^{(i)}$ is a matrix having null diagonal elements. The latter is estimated in each iteration by optimizing the first order Taylor expansion of $C_{ILS}$.\\[0.5cm] 
%
%------------------------------------------------------------
\underline{\bf LUCJD \cite{LUCJD}}: 
%------------------------------------------------------------
This algorithm is developed by Wang, Gong and Lin in 2012. It considers $C_{ILS}$ criterion as the CVFFDiag. It decomposes the mixing matrix in its LU form where $\textbf{L}$ and $\textbf{U}$ are lower and upper triangular matrices with diagonal entries equal to one.  This algorithm is developed in \cite{Afsari} for real NOJD and generalized to complex case in \cite{LUCJD}. Matrices $\textbf{L}$ and $\textbf{U}$ are optimized in alternating way by minimizing $C_{ILS}$. Note that the entries of $\textbf{L}$ and $\textbf{U}$ are updated one by one (keeping the other entries fixed).\\[0.5cm]
%
%------------------------------------------------------------
\underline{\bf ALS \cite{als}}: 
%------------------------------------------------------------
This algorithm is developed by Trainini and Moreau in 2011. It  minimizes $C_{DLS}$ criterion as the ACDC and relaxes the relationship between $\textbf{A}^H$ and $\textbf{A}$. The algorithm is developed by considering three steps, the first one estimates diagonal matrices $\textbf{D}_k$ by keeping $\textbf{A}$ and $\textbf{A}^H$ fixed. The second one uses the obtained $\textbf{D}_k$ and fixed $\textbf{A}^H$ to compute the mixing matrix $\textbf{A}$ and the last step uses the obtained $\textbf{D}_k$ and $\textbf{A}$ from the first and second steps, respectively, to estimate $\textbf{A}^H$. These steps are realized for each iteration and repeated until the convergence state is reached. \\%[0.5cm]
%
%\end{description}

Note that other algorithms exist in the literature, developed for special cases, but are not considered in our study. For example, in \cite{Pham} the developed algorithm is applied only for positive definite matrices. In \cite{DIEM}, the developed algorithm is a direct method (not iterative) which makes it more sensitive to difficult JD problem (the algorithm is not efficient when the number of matrices is less than the matrix dimension).

%------------------------------------------------------------
%------------------------------------------------------------
\section{Basic generalization of JDi algorithm} \label{G_JDi}
%------------------------------------------------------------
%------------------------------------------------------------
We introduce herein the basic generalization of JDi algorithm, given in \cite{JDi}, from real to complex case. First, the basic idea is to transform hermitian matrices obtained in (\ref{Mat_her}) to real symmetric ones given by (\ref{fonc}) to which, we apply the JDi algorithm. Then in section \ref{cjdi}, by modifying the first approach, we develop the CJDi algorithm which uses the hermitian matrices directly.%is applied directly to the hermitian matrices instead of real symmetric ones.

%------------------------------------------------------------
%------------------------------------------------------------
\subsection{Complex to real matrix transformation} \label{trans}
%------------------------------------------------------------
%------------------------------------------------------------
The first idea of our approach consists of  transforming the original  problem of complex matrix joint diagonalization into JD of real symmetric matrices which allows us to apply JDi algorithm. %From this solution, we obtained Generalized Joint Diagonalization (+) algorithms (GJDi, GJDi+).
Hence, we transform the $K$ complex matrices into $2K$ hermitian  matrices $\{\tilde{M}_k, ~ k=1, \cdots, 2K\}$ according to:

\begin{eqnarray}
\tilde{\textbf{M}}_{2k} & = &  \frac{1}{2}(\textbf{M}_{k}+\textbf{M}_{k}^H)=\textbf{A}\Re e(\textbf{D}_k)\textbf{A}^H \nonumber \\ 
 \tilde{\textbf{M}}_{2k-1} & = & \frac{1}{2{\jmath}}(\textbf{M}_{k}-\textbf{M}_{k}^H)= \textbf{A}\Im m(\textbf{D}_k)\textbf{A}^H.
\label{Mat_her}
\end{eqnarray}
where $\Re e()$ and $\Im m()$ refer to the real part and imaginary part of a complex entity, respectively. Now, the $N\times N$  hermitian  matrices are transformed into $2N\times 2N$ real matrices according to: 

\begin{eqnarray}
\label{fonc}
\textit{f} \left( \tilde{\textbf{M}}_k \right) &=& \mathcal{M}_k  \\
&=&  \left[
\begin{array}{ll}
\Re e\left(\tilde{\textbf{M}}_k\right) & \Im m\left(\tilde{\textbf{M}}_k\right) \\
-\Im m\left(\tilde{\textbf{M}}_k\right) & \Re e\left(\tilde{\textbf{M}}_k\right) 
\end{array}
\right] ,  k=1, \cdots, 2K \nonumber
\end{eqnarray}

Based on (\ref{Mat_her}) and (\ref{fonc}), one can easily see that matrices $ \mathcal{M}_k, ~  k=1, \cdots, 2K $  share the appropriate JD structure, i.e.
\begin{eqnarray}
\mathcal{M}_{2k}   & = & \mathcal{A} \left[ \begin{array}{ccc}\Re e({\textbf{D}}_k) & 0  \\ 0 & \Re e({\textbf{D}}_k) \end{array} \right] \mathcal{A}^T  \nonumber \\
\mathcal{M}_{2k-1} & = & \mathcal{A} \left[ \begin{array}{ccc}\Im m({\textbf{D}}_k) & 0  \\ 0 & \Im m({\textbf{D}}_k) \end{array} \right] \mathcal{A}^T \label{eqq5}
\end{eqnarray}
where $\mathcal{A} = \textit{f} \left( \textbf{A} \right)$. This property  allows us to apply JDi algorithm to achieve the desired joint diagonalization\footnote{Note that in (\ref{eqq5}), the diagonal entries of $\textbf{D}_k$ appear twice leading to an extra indeterminacy that should be taken into consideration when solving the complex JD problem (see lemma \ref{lemme_1} in subsection \ref{CDMR}).}.%(see section \ref{subsection4-2} for details) 

Like in the JDi method, the real diagonalizing matrix $\mathcal{V}$ associated to the complex one $\textbf{V}$, is decomposed as a product of generalized rotation matrices according to:
\begin{equation}
\mathcal{V} = \prod_{\# \mbox{sweeps}} ~ \prod_{ 1\leq i < j \leq 2 N} \mathcal{H}_{j}^{i}(\theta, y)
\end{equation}
where $ \# \mbox{sweeps} $ represents the sweeps (iterations) number and $\mathcal{H}_{j}^{i}(\theta, y)$ is the generalized rotation matrix given by:
\begin{equation}
\mathcal{H}_{j}^{i}(\theta, y) = \mathcal{S}_{j}^{i}(y)\mathcal{G}_{j}^{i}(\theta)
\end{equation}
$\mathcal{G}_{j}^{i}(\theta)$ and $\mathcal{S}_{j}^{i}(y)$ being the elementary Givens and Shear rotation matrices which are equal to the identity matrix except for their $(i,i)^{th}$,  $(i,j)^{th}$, $(j,i)^{th}$ , and $(j,j)^{th}$ entries given by:
\begin{equation}
\left[ \begin{array}{cc} \mathcal{G}^i_j(i,i) & \mathcal{G}^i_j(i,j)\\ \mathcal{G}^i_j(j,i) & \mathcal{G}^i_j(j,j) \end{array}\right]=\left[ \begin{array}{cc} \cos(\theta) & \sin(\theta)\\ -\sin(\theta) & \cos(\theta) \end{array}\right]
\end{equation}
\begin{equation}
\left[ \begin{array}{cc} \mathcal{S}^i_j(i,i) & \mathcal{S}^i_j(i,j)\\ \mathcal{S}^i_j(j,i) & \mathcal{S}^i_j(j,j) \end{array}\right]=\left[ \begin{array}{cc} \cosh(y) & \sinh(y)\\ \sinh(y) & \cosh(y) \end{array}\right]
\end{equation}
where $\theta$ and $y$ are the Givens angle and the Shear parameter, respectively. Based on these elementary transformations, we express next the transformed matrices as well as the JD criterion given in (\ref{cils}).  

%%%%%%%%%%%%%%%%%%%%%%%%%%%%%%%%%%%%%%%%%%%%%%%%%%%%%%%
 \subsection{Matrix transformations}
%%%%%%%%%%%%%%%%%%%%%%%%%%%%%%%%%%%%%%%%%%%%%%%%%%%%%%%

As shown in subsection \ref{trans}, the set of complex matrices is transformed into a set of real symmetric ones, $ \left\{ \mathcal{M}_k, ~k=1,...,2K\right\}$, to which all Givens and Shear rotations are applied. We denote by $ \mathcal{M}'_k,  ~k=1,\cdots,2K$ the updated matrices when using the elementary rotations, i.e.:
 \begin{equation}
 \mathcal{M}'_k=\mathcal{H}_{j}^{i}(\theta,y)\mathcal{M}_k \mathcal{H}_{j}^{i}(\theta,y)^T
 \label{update}
 \end{equation}
Note that only the $i^{th}$ and $j^{th}$ rows and columns of $\mathcal{M}_k$ are transformed so that $(i,j)^{th}$ entries are twice affected by the latter transformation. These entries can be expressed as:
\begin{equation}
\mathcal{M}'_k(i,j)=\textbf{v}^T\left[ \begin{array}{c}  \frac{\mathcal{M}_k(i,i)+\mathcal{M}_k(i,i)}{2} \\ \frac{\mathcal{M}_k(i,i)-\mathcal{M}_k(i,i)}{2} \\ \mathcal{M}_k(i,j) \end{array}\right] 
%\mathcal{M}'_k(i,j)=\left[ \begin{array}{ccc} \epsilon_{k,ij} & \delta_{k,ij} & \mathcal{M}_k(i,j) \end{array}\right] \textbf{v}%\begin{array}{ccc} \mathcal{M}'_k(i,j)  & = &   \epsilon_{k,ij} \sinh(2y) -\delta_{k,ij} \sin(2\theta)\cosh(2y)\\  &  & + ~ \mathcal{M}_k(i,j)\cos(2\theta)\cosh(2y) \end{array}
\label{matij}
\end{equation}
 
where
%\begin{equation}
%\left\{\begin{array}{lcr}
%\epsilon_{k,ij} &=& \frac{\mathcal{M}_k(i,i)+\mathcal{M}_k(j,j)}{2}\\
%\delta_{k,ij}  &=&  \frac{\mathcal{M}_k(i,i)-\mathcal{M}_k(j,j)}{2}
%\end{array}
%\right.
%\end{equation}
%and

\begin{equation}
\textbf{v} = \left[ \begin{array}{c} \sinh(2y)\\ -\sin(2\theta)\cosh(2y)\\ \cos(2\theta)\cosh(2y) \end{array}\right]
%\textbf{v}_2 = \left[ \begin{array}{c} \cosh(2y)\\ -\sin(2\theta)\sinh(2y)\\ \cos(2\theta)\sinh(2y) \end{array}\right]
\label{v1v2}
\end{equation}

Souloumiac in \cite{JDi} introduces a simplified JD criterion which is the sum of squares of $(i,j)^{th}$ entries. This JD criterion denoted $\mathcal{C}'_{ij}$ can be expressed by using (\ref{matij}) as:
\begin{equation}
%\begin{array}{lll}
\mathcal{C}'_{ij}  =  \sum_{k=1}^{2K}{\mathcal{M}'_k(i,j)^2} =  \textbf{v}^T \textbf{R} \textbf{v}
%\end{array}
\label{c1ij}
\end{equation}
where 
\begin{equation}
\textbf{R} = \textbf{W} \textbf{W}^T %,~~~~~~~~~~~~~~ \textbf{g}=\sum_{k=1}^{2K}\sum_{l\neq i,j}\textbf{g}_{kl}
\label{mat_R}
\end{equation}
and 
\begin{equation}
\textbf{W}=\left[ \begin{array}{ccc}  \frac{\mathcal{M}_1(i,i)+\mathcal{M}_1(i,i)}{2} & \cdots &  \frac{\mathcal{M}_{2K}(i,i)+\mathcal{M}_{2K}(i,i)}{2} \\ \frac{\mathcal{M}_1(i,i)-\mathcal{M}_1(i,i)}{2} & \cdots & \frac{\mathcal{M}_{2K}(i,i)-\mathcal{M}_{2K}(i,i)}{2} \\ \mathcal{M}_{1}(i,j) & \cdots & \mathcal{M}_{2K}(i,j) \end{array} \right]
\label{matw}
\end{equation} %the different parameters are given in (\ref{v1v2}) and (\ref{mat_R}). 
The results in \cite{JDi} show  that by minimizing this simplified criterion, joint diagonalization can be achieved in few iterations (see \cite{JDi} for more details).  %However, when the modulus of uniqueness is close to one, JDi algorithm diverges or becomes slowly convergent in which case the original criterion in (\ref{cij}) needs to be considered (see \cite{JDiplus} for more details).  

%%%%%%%%%%%%%%%%%%%%%%%%%%%%%%%%%%%%%%%%%%%%%%%%%%%%%%%
 \subsection{Direct generalization of JDi} %\label{al_jdi}
%%%%%%%%%%%%%%%%%%%%%%%%%%%%%%%%%%%%%%%%%%%%%%%%%%%%%%%

In JDi algorithm, $\mathcal{C}'_{ij}$ JD criterion given in (\ref{c1ij}) is minimized under the hyperbolic normalization as follows:
\begin{equation}
(\theta,y)=\mbox{arg}\min_{\textbf{v}^T \textbf{J} \textbf{v}=1} \left\{\textbf{v}^T \textbf{R} \textbf{v}  \right\}
\label{opt1}
\end{equation}
with $\textbf{J}=\left[ \begin{array}{ccc} -1 & 0 &0 \\ ~~0& 1&0 \\ ~~0 & 0 &1 \end{array} \right] $. \\
 %In the improved version JDi+ given in \cite{JDiplus}, the original JD criterion, $\mathcal{C}_{ij}$,  is minimized using numerical methods. The minimization process is initialized by the JDi solution then improved by one or few Newton iterations taking $\mathcal{C}_{ij}$ given in equation (\ref{cij}) as a non linear cost function of $(\theta, y)$. 
%%%%%%%%%%%%%%%%%%%%%%%%%%%%%%%%%%%%%%%%%%%%%%%%%%%%%%%
% \subsubsection{Basic JDi algorithm} \label{al_jdi}
%%%%%%%%%%%%%%%%%%%%%%%%%%%%%%%%%%%%%%%%%%%%%%%%%%%%%%%
%As given in \cite{JDi}, optimal parameters $(\theta,y)$ can be found by minimizing the quadratic form given in (\ref{c1ij}) under the quadratic constrain $\textbf{v}^T\textbf{J}\textbf{v}=1$. 
The solution of (\ref{opt1}) is the eigenvector associated to the median generalized eigenvalue of $(\textbf{R},\textbf{J})$ denoted as $\textbf{v}=\begin{array}{ccc} \left[\textit{v}_1\right. & \textit{v}_2& \left.\textit{v}_3\right]^T \end{array}$. \\
Then, the optimal parameters can be expressed as :
\begin{equation}
\left\{\begin{array}{ccc}
\cosh(y) & = & \frac{1}{\sqrt{2}}\sqrt{1+\sqrt{1+\textit{v}_1^2}} \\ 
 & & \\
\sinh(y) & = & \frac{\textit{v}_1}{2\cosh{y}} \label{ytheta}\\
 & & \\
\cos(\theta) & = & \frac{1}{\sqrt{2}}\sqrt{1+\frac{\textit{v}_3}{\sqrt{1+\textit{v}_1^2}}} \\ 
 & & \\
\sin(\theta) & = & -\frac{\textit{v}_2}{2\cos{\theta} \sqrt{1+\textit{v}_1^2}} 
\end{array}\right.
\end{equation}
\\A normalization is introduced in JDi algorithm. It ensures that the estimated mixing matrix has columns of equal norm and determinant equal to one. %This step guarantees the algorithm's stability and avoids divergence that might be caused by the non orthogonal Shear rotations. \\ 
The normalizing matrix $\mathcal{D}_a$ can be expressed as: % \cite{JDi} 

\begin{equation}
\mathcal{D}_a=\frac{1}{\sqrt[2N]{ \left\|\textbf{a}_1\right\| \cdots \left\|\textbf{a}_{2N}\right\|}}\left[  \begin{array}{ccc} \frac{1}{\left\|\textbf{a}_1\right\|}\ & & 0 \\ & \ddots & \\ 0& &\frac{1}{\left\|\textbf{a}_{2N}\right\|} \end{array} \right]
\label{mat_da}
\end{equation}
where $\textbf{a}_1, ..., \textbf{a}_{2N}$ are the columns of the estimated mixing matrix $\tilde{\mathcal{A}}$. \\We propose here to modify the JDi algorithm by  replacing the previous normalization step by the following eigenvector normalization: 
\begin{equation}
\textbf{v}\leftarrow \frac{\textbf{v}}{\sqrt{\left|\textbf{v}^T \textbf{J} \textbf{v}\right|}}
\label{Vnor}
\end{equation}
The modified JDi algorithm is summarized in Table \ref{JDi_algo}, where $\tau$ and $M_{it}$ are a fixed threshold and maximum sweep number respectively, chosen to stop the algorithm. %The improved version JDi+ has the same structure but it introduces a numerical optimization just after computing JDi solution and before updating the symmetrical matrices as shown in the same Table. %This numerical optimization is done by minimizing the non linear function given in (\ref{cij}) some Newton iteration 
\begin{table}[tb]
\centering
\begin{tabular}{|*{1}l |}

\hline 
\\
\textbf{Require} : $ \mathcal{M}_k,~k=1,\cdots,2K$, a fixed threshold $\tau$\\  ~~~~~~~~~and a maximum sweep number $M_{it}$. \\
\textbf{Initialization}:
~$\tilde{\mathcal{V}}=I_{2N}$.\\%, $\tilde{\mathcal{A}}=I_{2N}$
~\textbf{while} $\max_{i,j}(\left|\sinh(y)\right|,\left|\sin(\theta)\right|)>\tau$ and (\#sweeps $< M_{it}$)\\
~~~~\textbf{for} all $1 \leq i <j \leq 2N$\\

~~~~~~Build $\textbf{R}$ as in (\ref{mat_R}).\\

%~~~~~~Compute the eigenvector $\textbf{v}=\left[ \begin{array}{ccc} \textit{v}_1 & \textit{v}_2 & \textit{v}_3  \end{array}\right]^T$ \\
~~~~~~Compute the solution $\textbf{v}$ of (\ref{opt1}).\\

~~~~~~Normalize vector  $\textbf{v}$ as in (\ref{Vnor}). \\

~~~~~~if $(\textit{v}_3<0)$   then  $(\textbf{v} \leftarrow -\textbf{v})$.\\

~~~~~~Compute $\cos(\theta), \sin(\theta), \cosh(y)$ and $\sinh(y)$ as in (\ref{ytheta}). \\

%~~~~~~Correct ($\theta , y$) by using some Newton iterations\footnotemark. \\

~~~~~~Update matrices $\mathcal{M}_k$ as in (\ref{update})\\
~~~~~~~  $\tilde{\mathcal{V}}\leftarrow \mathcal{H}_{j}^{i}(\theta,y)\tilde{\mathcal{V}}$.\\
%~~~~~~~~~~ $\tilde{\mathcal{A}}\leftarrow \tilde{\mathcal{A}}\mathcal{H}_{j}^{i}(\theta,y)^{-1}$\\

~~~~\textbf{end for} \\
%~~\textbf{Normalization step}\\
%
%~~~~Compute $\mathcal{D}_a$ as in (\ref{mat_da}) \\
%~~~~$\tilde{\mathcal{A}}\leftarrow \tilde{\mathcal{A}} \mathcal{D}_a$\\
%~~~~$\tilde{\mathcal{V}}\leftarrow \mathcal{D}_a^{-1} \tilde{\mathcal{V}} $\\
%~~~~$\mathcal{M}_k\leftarrow \mathcal{D}_a^{-1}\mathcal{M}_k \mathcal{D}_a^{-1} $ for $k=1, \cdots, 2K$.\\
     
~\textbf{end while}\\
\\
 \hline
\end{tabular}
\caption{Modified JDi algorithm} 
\label{JDi_algo}
\end{table}
%\footnotetext{This step is introduced only in JDi+ algorithm \cite{JDiplus}.}
\\At this stage, we only applied Modified JDi algorithm to the set of transformed real symmetric matrices $\mathcal{M}_k$. Once the algorithm converges, we get $\tilde{\mathcal{V}}$ (equivalently $\tilde{\mathcal{A}}^{-1}=\tilde{\mathcal{V}}$) up to some inherent indeterminacies. Now, the question is how to get the complex diagonalizing matrix $\tilde{\textbf{V}}$ and get rid of the undesired indeterminacies. This question is considered in the next subsection.  

%%%%%%%%%%%%%%%%%%%%%%%%%%%%%%%%%%%%%%%%%%%%%%%%%%%%%%%
 \subsection{Complex diagonalizing matrix retrieval} \label{CDMR}
%%%%%%%%%%%%%%%%%%%%%%%%%%%%%%%%%%%%%%%%%%%%%%%%%%%%%%%

As mentioned in section \ref{prob}, the JD problem has inherent indeterminacies in the sense that matrix $\textbf{V}$ is estimated up to permutation and diagonal matrices. However, the specific structure of our real valued matrices $\left\{\mathcal{M}_{\left.k\right|k=1...2K}\right\}$ given in (\ref{fonc}) leads to extra indeterminacies according to the following lemma:
\newtheorem{theoreme}{Lemma}
\begin{theoreme}
Define vectors $\textbf{d}_{i}$, $i=1, ... ,N$ as $$ \textbf{d}_i=  \begin{array}{lcr} \left[D_1\left(i,i\right)\right. & ... & \left.D_K\left(i,i\right)\right] \end{array} $$
Under the condition that the $1 \times 2K$ dimensional vectors $\left[  \Re e \left(\textbf{d}_i\right),\right.$ $\left.\Im m \left(\textbf{d}_i\right)  \right]$
are pairwise linearly independent, the JD problem's solution of $\left\{\mathcal{M}_1,...,\mathcal{M}_{2K}\right\}$ is such that, there exists a permutation matrix $\textbf{P}$ satisfying:
$$ \tilde{\mathcal{A}} \textbf{P} = \left[ \mathcal{A}_1 , ... ,\mathcal{A}_N\right]$$ 
with $\mathcal{A}_i=\left[ \mathcal{A}(:,i),\mathcal{A}(:,i+N)\right] \textbf{U}_i$, $\mathcal{A}(:,i)$ being the $i^{th}$ column vector of $\mathcal{A}$ and $\textbf{U}_i$ is a $2\times 2$ orthogonal matrix i.e. $\textbf{U}_i \textbf{U}_i^{T}=\alpha_i \textbf{I}_2$ for a given scalar factor $\alpha_i$.

\label{lemme_1}
\end{theoreme}
\textbf{Proof}: 
This result can be deduced directly from Theorem 3 in \cite{IDA2}.\\

To retrieve the original complex matrix $\textbf{A}$ from the estimated matrix $\tilde{\mathcal{A}}$, one needs to find the permutation that associates correctly the $i^{th}$ column of $\mathcal{A}$ to its $\left(i+N\right)^{th}$ one.
For that, since matrix $\textbf{U}_i$ is orthogonal, one can represent it as:
\begin{equation}
 \textbf{U}_i= \left[\begin{array}{cc} a_i & b_i \\ -b_i & a_i \end{array} \right]
\label{Ui}
\end{equation}
and hence
\begin{equation}
\begin{array}{ll}
 \mathcal{A}_i & = \left[\begin{array}{cc} \Re e (\textbf{a}_i) & \Im m(\textbf{a}_i) \\ - \Im m(\textbf{a}_i) & \Re e(\textbf{a}_i) \end{array} \right] \left[\begin{array}{cc} a_i & b_i \\ -b_i & a_i \end{array} \right]\\ & \\
 & = \left[\begin{array}{cc} \Re e (\tilde{\textbf{a}}_i) & \Im m(\tilde{\textbf{a}}_i) \\ - \Im m(\tilde{\textbf{a}}_i) & \Re e(\tilde{\textbf{a}}_i) \end{array} \right]
 \end{array}
\label{AAi}
\end{equation}
where $\textbf{a}_i$ is the $i^{th}$ column vector of $\textbf{A}$ and $\tilde{\textbf{a}}_i=c_i \textbf{a}_i $ with $c_i=a_i + \jmath b_i$ ($\jmath=\sqrt{-1}$) and $a_i$, $b_i$ are given in (\ref{Ui}).\\
From (\ref{AAi}), one can observe that two columns of $\mathcal{A}$:\\
\begin{equation*}
\begin{array}{ccc}
\tilde{\textbf{a}}_i=\left[ \begin{array}{c} \tilde{\textbf{a}}_{i,1}\\ \tilde{\textbf{a}}_{i,2} \end{array}\right] & \mbox{and} &  \tilde{\textbf{a}}_j=\left[ \begin{array}{c} \tilde{\textbf{a}}_{j,1}\\ \tilde{\textbf{a}}_{j,2} \end{array}\right]
\end{array}
\end{equation*}
can be associated if they satisfy the relation :
\begin{equation}
\left\{ \begin{array}{cc} \tilde{\textbf{a}}_{i,1}-\tilde{\textbf{a}}_{j,2}& =0\\ \tilde{\textbf{a}}_{i,2}+\tilde{\textbf{a}}_{j,1}& =0 \end{array}\right.
\label{asso}
\end{equation}
In practice, we solve equation (\ref{asso}) in the least squares sense to take into account the estimation errors.
Once, this pairing process is achieved, the $i^{th}$ column of matrix $\textbf{A}$ is estimated (up to a scalar complex valued factor) from the first column of matrix $\mathcal{A}_i$ as $\tilde{\textbf{a}}_{i}=\tilde{\textbf{a}}_{i,1}-\jmath \tilde{\textbf{a}}_{i,2}$.\\
Similarly, if $\textbf{P}$ is the permutation pairing correctly the columns of $\tilde{\mathcal{A}}$, then $\textbf{P}^T$ is the one pairing correctly the rows of $\tilde{\mathcal{V}}$: i.e. $\tilde{\mathcal{V}} \leftarrow \textbf{P}^T\tilde{\mathcal{V}} \ $.\\[0.2cm]
 
\begin{table}[tb]
\centering
\begin{tabular}{|*{1}l |}
\hline 
\\
\textbf{Step 1}: Transform the complex matrices into real symmetric ones \\%
~~~~~~~~~~~~~~~as explained in subsection \ref{trans}. \\
%~~~~~~~~~~~~~~~Use equation (\ref{Mat_her}) to get hermitian matrices; \\
%~~~~~~~~~~~~~~~Apply (\ref{fonc}) to obtain real symmetric matrices as in (\ref{eqq5}).\\
\\
\textbf{Step 2}: Apply the modified JDi algorithm as given in Table \ref{JDi_algo}.\\
\\
\textbf{Step 3}: Retrieve the complex diagonalizing matrix \\%As
~~~~~~~~~~~~~~~as developed in subsection \ref{CDMR}. \\
\\
 \hline
\end{tabular}
\caption{Basic JDi algorithm generalization} \label{table1_1}
\end{table}
%

%These first developments  are summarized in Table \ref{table1_1} as a basic generalization of JDi algorithm to the complex case. %Note that 
 
%%%%%%%%%%%%%%%%%%%%%%%%%%%%%%%%%%%%%%%%%%%%%%%%%%%%%%%
 \section{CJDi algorithm} \label{cjdi}
%%%%%%%%%%%%%%%%%%%%%%%%%%%%%%%%%%%%%%%%%%%%%%%%%%%%%%%
In this section, we give, first, the algorithm's development based on real symmetric matrices given in equation (\ref{fonc}). Then, direct complex implementation is developed.\\% which is equivalent to the first one. In the latter implementation, the algorithm is applied to hermitian matrices where we applied complex Shear and Givens rotations.\\
 
%------------------------------------------------------
\subsection{Algorithm's development based on real matrices}
%------------------------------------------------------

Note that the real symmetric matrices given in (\ref{fonc}) have a special structure which can be used to simplify and improve the previous generalization of JDi algorithm. More precisely, we look for transformations that preserve the considered matrix structure along the iterations  which allows us to skip the step of complex diagonalizing (resp. mixing) matrix retrieval. Indeed, in the basic generalization,  the introduction of the elementary rotation matrix $\mathcal{H}_{j}^{i}(\theta,y)$ causes the loss of matrix structure. For example, when rotation indices are $1 \leq i < j \leq N$  then the updated entries $\mathcal{M}_k(i,i)$ and $\mathcal{M}_k(j,j)$ are not anymore equal to $\mathcal{M}_k(i+N,i+N)$ and $\mathcal{M}_k(j+N,j+N)$, respectively. Hence, to preserve the considered matrix structure, one needs to introduce a second elementary rotation which is $\mathcal{H}^{i+N}_{j+N}(\theta,y)$. The following lemma provides the solution to the previous problem.

\begin{theoreme} 

%\begin{enumerate}
%	\item 
1) If the  real symmetric matrices $\left\{\mathcal{M}_k\right\}$ are updated by the elementary rotation matrix $\mathcal{H}^{i}_{j}(\theta,y)$, $1 \leq i < j \leq N$, then the second elementary rotation which preserves the matrix structure in (\ref{fonc}) is $\mathcal{H}^{i+N}_{j+N}(\theta,y)$, i.e. the generalized rotation matrix with same angle and Shear parameters. \\

2) If the  real symmetric matrices $\left\{\mathcal{M}_k\right\}$ are updated by the elementary rotation matrix $\mathcal{H}^{i}_{j+N}(\theta,y)$, $1 \leq i < j \leq N$, then the second elementary rotation which preserves matrix structure in (\ref{fonc}) is $\mathcal{H}^{j}_{i+N}(\theta,-y)$ where the sign of the Shear parameter is inverted. 
\label{lemme1}
\end{theoreme}

\noindent\textbf{Proof}: The proof of this lemma is given in appendix \ref{appendix_A}. \\

The rotation parameters are now optimized in such a way we minimize the simplified JD criterion given in (\ref{c1ij}) for the transformed matrices:
\begin{equation}
\mathcal{M}''_k=\mathcal{H}^{i+N}_{j+N}(\theta,y)\mathcal{H}^i_j(\theta,y)\mathcal{M}_k\mathcal{H}^i_j(\theta,y)^T\mathcal{H}^{i+N}_{j+N}(\theta,y)^T
\label{trans_1}
\end{equation}
and
\begin{equation}
\mathcal{M}'''_k=\mathcal{H}^{j}_{i+N}(\theta,-y)\mathcal{H}^{i}_{j+N}(\theta,y)\mathcal{M}_k\mathcal{H}^{i}_{j+N}(\theta,y)^T\mathcal{H}^{j}_{i+N}(\theta,-y)^T
\label{trans_2}
\end{equation}
Interestingly, the estimation of the optimal parameters is the same as the one we obtained before when using the matrix transformation in (\ref{update}).

\begin{theoreme}
The JD criterion $C'_{ij}$ calculated with the matrix transform (\ref{update}) for indices ($i$,$j$) (resp. the matrix transform (\ref{update}) for indices ($i$,$j+N$)) is equal, up to constant factors, to the one calculated with the matrix transform (\ref{trans_1})  (resp. the matrix transform (\ref{trans_2})), i.e.
\begin{equation}
\begin{array}{ll} \sum_{k=1}^{2K} \left[{\mathcal{M}''_k(i,j)^2+\mathcal{M}''_k(i+N,j+N)^2}\right. & \\ \left.+ {\mathcal{M}''_k(i,j+N)^2+\mathcal{M}''_k(j,i+N)^2}\right]& \\ =2\sum_{k=1}^{2K}{\mathcal{M}'_k(i,j)^2}+c_1 & \end{array} 
\label{ciNjN}
\end{equation}

\begin{equation} \begin{array}{ll} \sum_{k=1}^{2K} \left[{\mathcal{M}'''_k(i,j)^2+\mathcal{M}'''_k(i+N,j+N)^2}\right. & \\ \left.+ {\mathcal{M}'''_k(i,j+N)^2+\mathcal{M}'''_k(j,i+N)^2}\right]& \\=2\sum_{k=1}^{2K}{\mathcal{M}'_k(i,j+N)^2}+c_2 &  \end{array} 
\label{cijN}
\end{equation}
where $c_1$ and $c_2$ are scalar constants independent from the angle and shear parameters.
Consequently, the  optimal parameters obtained by minimizing (\ref{ciNjN}) with indices ($i$, $j$) (resp.  (\ref{cijN}) with indices ($i$, $j+N$) ) are the same as the one obtained by minimizing (\ref{c1ij}) with indices ($i$, $j$) (resp. (\ref{c1ij}) with indices ($i$, $j+N$)).
\label{lemme3}
\end{theoreme}
\textbf{Proof}: The proof is given in appendix \ref{appendix_B}.\\[0.1cm]

Compared to the previous basic generalization of JDi algorithm, the developed algorithm called CJDi  preserves our matrix structure and decreases the number of iterations per sweep. In the first generalization, the number of iterations (index pairs) per sweep is $(2N-1)N$ while in CJDi, this number   decreases to $N\left(N-1\right)$. Also, CJDi takes into account some extra information about the matrix structure which leads to a slight performance improvement.

\begin{table}[tb]
\centering
\begin{tabular}{|*{1}l |}
\hline 
\\ 
\textbf{Require} : $ \textbf{M}_k,~k=1,\cdots,K$, a fixed threshold $\tau$\\ ~~~~~~~~~and a maximum sweep number $M_{it}$. \\
\textbf{Initialization}:
~ $\tilde{\textbf{V}}=I_{N}+\jmath I_{N}$.\\ %and $\tilde{\textbf{A}}=I_{N}+\jmath I_{N}$
~~~Compute $\left\{\tilde{\textbf{M}}_k|k=1,\cdots ,2K\right\}$ as in (\ref{Mat_her}).   \\
~\textbf{while} $\max_{i,j}(\left|\sinh(y)\right|,\left|\sin(\theta)\right|)>\tau$
 and (\#sweeps $< M_{it}$)\\
~~~~\textbf{for} all $1 \leq i <j \leq N$\\
%~~~~~~~\textbf{for} $j =i+1,\cdots, N$\\
~~~~~~~~~~ Estimate $ \textbf{H}_{1,ij}(\theta,y)$ using (\ref{mij1c}) and (\ref{opt1}). \\
~~~~~~~~~~ Update matrices $\left\{\tilde{\textbf{M}}_k|k=1,\cdots ,2K\right\}$ as in (\ref{trans_1c}).\\
~~~~~~~~~~$ \tilde{\textbf{V}} \leftarrow \textbf{H}_{1,ij}(\theta,y)  \tilde{\textbf{V}}$ \\
%~~~~~~~~~~$ \tilde{\textbf{A}} \leftarrow \tilde{\textbf{A}} \textbf{H}_{1,ij}(\theta,y)^{-1}$ \\
\\
~~~~~~~~~~ Estimate $ \textbf{H}_{2,ij}(\theta,y)$ using (\ref{mij2c}) and (\ref{opt1}).\\
~~~~~~~~~~ Update matrices $\left\{\tilde{\textbf{M}}_k|k=1,\cdots ,2K\right\}$ as in (\ref{trans_2c}).\\
~~~~~~~~~~$ \tilde{\textbf{V}} \leftarrow \textbf{H}_{2,ij}(\theta,y) \tilde{\textbf{V}}$\\
%~~~~~~~~~~$ \tilde{\textbf{A}} \leftarrow \tilde{\textbf{A}} \textbf{H}_{2,ij}(\theta,y)^{-1}$\\
~~~~\textbf{end for}\\
\textbf{end while}.\\ 
\hline
\end{tabular}
\caption{CJDi algorithm } \label{algo2}
\end{table}

%------------------------------------------------------------	
\subsection{Complex implementation } \label{eff}
%------------------------------------------------------------
The matrix rotations in (\ref{trans_1}) and (\ref{trans_2}) can be rewritten in the complex form by reversing the function in (\ref{fonc}), i.e. using the real to complex transformation $\textbf{M}_k=f^{-1}\left(\mathcal{M}_k\right)$. Considering this transformation, one can express the previous rotations as:
%Observing the previous rotation structure, the product $\mathcal{H}^{i+N}_{j+N}(\theta,y)$ $\mathcal{H}^i_j(\theta,y)$ can be obtained by  transforming  $\textbf{H}_{1,ij}(\theta,y)$ to the real symmetric once using equation (\ref{fonc}). The other rotation product $\mathcal{H}^{i}_{j+N}(\theta,y)$ $\mathcal{H}^j_{i+N}(\theta,y)$ can be also obtained by transforming the $\textbf{H}_{2,ij}(\theta,y)$ to the real symmetric once using the same equation (\ref{fonc}). where  $\textbf{H}_{1,ij}(\theta,y)$ and  $\textbf{H}_{2,ij}(\theta,y)$ can be expressed as:
\begin{eqnarray}
\textbf{H}_{1,ij}(\theta,y)& = & f^{-1}\left(\mathcal{H}^i_j\left(\theta,y\right) \mathcal{H}^{i+N}_{j+N}\left(\theta,y\right) \right) \nonumber \\
& = & \textbf{S}_{ij}(0,y) \textbf{G}_{ij}(0,\theta)
\label{Hc1}
\end{eqnarray}

\begin{eqnarray}
\textbf{H}_{2,ij}(\theta,y)& = & f^{-1}\left(\mathcal{H}^i_{j+N}\left(\theta,y\right) \mathcal{H}^{j}_{i+N}\left(\theta,-y\right) \right) \nonumber \\
& = & \textbf{S}_{ij}(\frac{\pi}{2},y) \textbf{G}_{ij}(\frac{\pi}{2},\theta)
\label{Hc2}
\end{eqnarray}
where $\textbf{S}_{ij}(\phi,y)$ $ \textbf{G}_{ij}(\alpha,\theta)$ are elementary complex Shear and Givens rotations which are equal to the identity matrix except for $(i,i)^{th}$, $(i,j)^{th}$, $(j,i)^{th}$ and $(j,j)^{th}$ entries given by:
\begin{equation}
\begin{array}{ll}
\left[ \begin{array}{ll} S_{ij}(i,i) & S_{ij}(i,j) \\ S_{ij}(j,i) & S_{ij}(j,j) \end{array} \right] = & \left[ \begin{array}{ll} \cosh(y) & e^{\jmath \phi} \sinh(y) \\ e^{-\jmath \phi} \sinh(y) & \cosh(y) \end{array} \right]
\end{array}
\label{shear}
\end{equation}%

\begin{equation}
\begin{array}{cc}
\left[ \begin{array}{ll} G_{ij}(i,i) & G_{ij}(i,j) \\ G_{ij}(j,i) & G_{ij}(j,j) \end{array} \right] = & \left[ \begin{array}{cc} \cos(\theta) & e^{\jmath \alpha} \sin(\theta) \\ -e^{-\jmath \alpha} \sin(\theta) & \cos(\theta) \end{array} \right]
\end{array}
\label{givens}
\end{equation}
Equations (\ref{trans_1}) and (\ref{trans_2}) correspond to the complex matrix transformations:%Note that all matrices preserve the structure given in (\ref{fonc}) hence it is better to work with hermitian matrices than with real once. The rotation parameters are now optimized in such a way we minimize the simplified JD criterion given in (\ref{c1ij}) for the hermitian matrices like:
\begin{equation}
\tilde{\textbf{M}}'_k=\textbf{H}_{1,ij}(\theta,y)\tilde{\textbf{M}}_k\textbf{H}_{1,ij}(\theta,y)^H
\label{trans_1c}
\end{equation}

\begin{equation}
\tilde{\textbf{M}}''_k=\textbf{H}_{2,ij}(\theta,y)\tilde{\textbf{M}}_k\textbf{H}_{2,ij}(\theta,y)^H
\label{trans_2c}
\end{equation}
The CJDi algorithm is summarized in Table \ref{algo2}.\\[0.5cm] 
\textbf{Remark:}
Using the previous equations (\ref{trans_1c}) and (\ref{trans_2c}), one can express the $(i,j)^{th}$ entries  as:

\begin{equation}
\tilde{\textbf{M}}'_k(i,j)=  \textbf{v}^T\left[ \begin{array}{c} \frac{\tilde{\textbf{M}}_k(i,i)+\tilde{\textbf{M}}_k(j,j)}{2} \\ \frac{\tilde{\textbf{M}}_k(i,i)-\tilde{\textbf{M}}_k(j,j)}{2} \\ \Re e(\tilde{\textbf{M}}_k(i,j)) \end{array}\right]  +\jmath \Im m(\tilde{\textbf{M}}_k(i,j))
\label{mij1c}
\end{equation}

\begin{equation}
\tilde{\textbf{M}}''_k(i,j)=  \Re e(\tilde{\textbf{M}}_k(i,j)) +\jmath \textbf{v}^T\left[ \begin{array}{c} \frac{\tilde{\textbf{M}}_k(i,i)+\tilde{\textbf{M}}_k(j,j)}{2} \\ \frac{\tilde{\textbf{M}}_k(i,i)-\tilde{\textbf{M}}_k(j,j)}{2} \\ \Im m(\tilde{\textbf{M}}_k(i,j)) \end{array}\right]  
\label{mij2c}
\end{equation}
Hence, applying transformation (\ref{trans_1c}) (resp. (\ref{trans_2c})) updates only the real part (resp. the imaginary part) of the $(i,j)^{th}$ entries.

%%-----------------------------------------------------
\subsection{Computational cost} \label{cost}
%%-----------------------------------------------------
We provide here an evaluation of the computational cost of our algorithm expressed in terms of real flops number (i.e. real multiplication plus real addition) per iteration (sweep).\\
In our evaluation, we took into account the matrices symmetry and the particular structure of the transformation matrices in (\ref{Hc1}) and (\ref{Hc2}) which entries are either real or pure imaginary numbers. Tacking this into consideration, the matrices product in (\ref{trans_1c}) or (\ref{trans_2c}) cost for each matrix $\tilde{\textbf{M}}_k$, $8N^2+O(N)$ flops (instead of $32N^2$ for a brute force implementation).\\
This numerical cost evaluation is performed similarly for the other considered NOJD algorithms and summarized in Table \ref{algo_com}.

\begin{table}[tb]
\centering
\begin{tabular}{|l|c|}
\hline
Algorithm & Number of real flops per sweep \\
\hline
CJDi & $16 K N^3$ + $\mathcal{O}\left(KN^2\right)$ \\
\hline
ACDC & $20 K N^3$+ $\mathcal{O}\left(N^4\right)$+$\mathcal{O}\left(N^2\right)$ \\
\hline
FAJD & $12 K N^4$ + $\mathcal{O}\left(N^3\right)$ \\
\hline
QDiag & $36 K N^3$ + $\mathcal{O}\left(KN^2\right)$ \\
\hline
UWEDGE & $8 K N^3$ + $\mathcal{O}\left(KN^2\right)$ \\
\hline
JUST & $32 K N^3$ + $\mathcal{O}\left(KN^2\right)$ \\
\hline
CVFFDiag & $8 K N^3$ + $\mathcal{O}\left(KN^2\right)$ \\
\hline
LUCJD & $24 K N^3$ + $\mathcal{O}\left(KN^2\right)$ \\
\hline
ALS & $20 K N^3$ + $\mathcal{O}\left(N^4\right)$+$\mathcal{O}\left(N^2\right)$ \\
\hline
\end{tabular}
\caption{NOJD Algorithms complexity} \label{algo_com}
\end{table}

%%-----------------------------------------------------
\section{Comparative performance analysis} \label{simul}
%%-----------------------------------------------------
The aim of this section is to compare CJDi with the different existing NOJD algorithms, cited in section \ref{NOJD}, for different scenarios. More precisely, we have chosen to evaluate and compare the algorithms sensitiveness to different factors that may affect the JD quality. These different criteria of comparison used in our study, are described in the next subsection.

%%-----------------------------------------------------
\subsection{Factors affecting NOJD problem}
%%-----------------------------------------------------

In 'good JD conditions' most algorithms  perform well, with slight differences in terms of convergence rate or JD quality. However, in adverse JD conditions, many of these algorithms lose their effectiveness or otherwise diverge. In this study, adverse conditions are accessible through the following factors: 
\begin{itemize}
\item \textit{Modulus of uniqueness} (MOU): defined in \cite{Afs}\cite{JDi} as the maximum correlation factor of vectors $\textbf{d}_i=\left[D_1(i,i),...\right.$ $\left.,D_K(i,i)\right]^T$, i.e.
\begin{equation}
\mbox{MOU}=\max_{i,j} \left(\frac{\left|\textbf{d}_i^H \textbf{d}_j\right|}{\left\| \textbf{d}_i\right\| \left\| \textbf{d}_j\right\| }\right)
\label{MOU}
\end{equation}
It is shown in \cite{JDi}\cite{JDiplus} that the JD quality decreases when the MOU get close to 1. 

\item \textit{Mixing matrix condition number}: The JD quality depends on the good or ill conditioning of mixing matrix $\textbf{A}$ (denoted Cond(\textbf{A})). The comparative study reveals the algorithm's robustness w.r.t. Cond($\textbf{A}$).

\item \textit{Diagonal matrices condition number}: In BSS problem, the power range of source signals affects the conditioning of diagonal matrices $\textbf{D}_k$ and hence the source separation quality. By comparing the algorithms sensitiveness w.r.t. this factor, we reveal their potential performance if used for separating sources with high dynamical range.

\item \textit{Matrix dimensions}: The JD problem is more difficult for large dimensional  matrices ($N>>1$) and hence we compare the algorithms performance in the following two cases: $N=5$ (small dimension) and $N=50$ (large dimension).
\item \textit{Noise effect}: In practice, matrices $\textbf{M}_k$ are given by some sample averaged statistics and hence are affected by finite sample size and noise effects. In that case, the exact JD (EJD) becomes approximate JD (AJD) and the algorithms performance is lower bounded by the noise level as shown by our comparative study. The algorithms behaviour and robustness w.r.t. noise effect is investigated in our study.
\end{itemize}

%------------------------------------------------------------

%%-----------------------------------------------------
\subsection{Simulation experiment set up}
%%-----------------------------------------------------

First, we have chosen the classical performance index ($PI$) to measure the JD quality. It can be expressed as \cite{Pindice}:
\begin{equation}
\begin{array}{cc}
PI\left(\textbf{G}\right)=&\frac{1}{2N\left(N-1\right)}\sum_{n=1}^{N}\left( \sum_{m=1}^{N}{\frac{\left|G(n,m)\right|^{2}}{\max_k \left|G(n,k)\right|^{2}}} -1\right)\\ & + \\ & \frac{1}{2N\left(N-1\right)}\sum_{n=1}^{N}\left( \sum_{m=1}^{N}{\frac{\left|G(m,n)\right|^{2}}{\max_k \left|G(k,n)\right|^{2}}} -1\right)
\end{array}
\end{equation}
where $G(n,m)$ is the ($n,m$)$^{th}$ entry of global matrix $\textbf{G}=\hat{\textbf{V}}\textbf{A}$, $\textbf{A}$ being the generated mixing matrix and $\hat{\textbf{V}}$ the estimated diagonalizing matrix. The closer the $PI$ is to zero, the better is the JD quality. Contrary to other existing criteria, the $PI$ is common to  all algorithms and allows us to compare them properly. \\

\begin{figure}[htbp]
		\begin{center}
	  \includegraphics[scale=0.4]{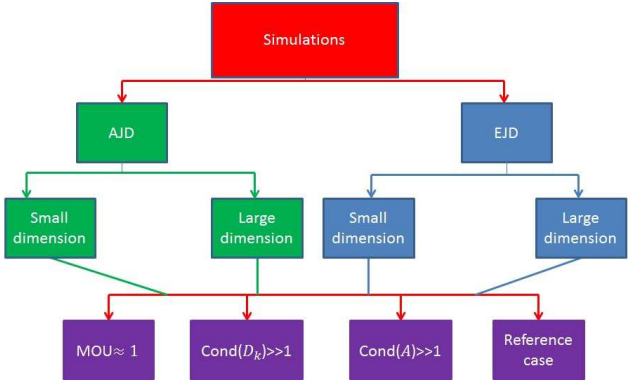}  
	  \caption{Simulation scenarios.}
	  \label{schema}
		\end{center}       
\end{figure}

Figure \ref{schema} illustrates the organization of our simulation scenarios where simulations are divided in two sets. The first one is dedicated to the exact joint diagonalization (EJD) case whilst the second one assess the algorithms performance in the AJD case by investigating the noise effect as given by equation (\ref{eq2}). Also, for each experiment, we  have considered two scenarios namely the small dimensional case ($N=5$) and the large dimensional case ($N=50$).\\
Finally, for each of these cases, we run four simulation experiments: (\textbf{i}) a reference simulation of relatively good conditions where MOU$\mbox{ }<0.6$ (resp. MOU$\mbox{ }\approx 0.9$ for large dimension), Cond($\textbf{A}$)$\mbox{ }<5$ (resp. Cond($\textbf{A}$)$\mbox{ }<50$ for large dimension) and Cond($\textbf{D}_k$)$\mbox{ }<10$ (resp. Cond($\textbf{D}_k$)$\mbox{ }<50$ for large dimension); (\textbf{ii}) a simulation experiment with MOU$\mbox{ }>1-10^{-6}$; (\textbf{iii}) a simulation experiment with Cond($\textbf{A}$)$\mbox{ }>100$ and (\textbf{iv}) a simulation experiment with Cond($\textbf{D}_k$)$\mbox{ }>10^{4}$.\\

%%--schéme de simulations ----------------------------------
%%-----------------------------------------------------

The other simulation parameters are as follows: mixing matrix entries are generated as independent and normally distributed variables of unit variance and zero mean. Similarly, the diagonal entries of $\textbf{D}_k$ are independent and normally distributed variables of unit variance and zero mean except in the context where Cond($\textbf{D}_k$)$\mbox{ }>10^4$ in which case $D_k(2,2)$ has a standard deviation of $10^{-4}$ or in the context where MOU$\mbox{ }>1-10^{-6}$ in which case $D_k(2,2)=D_k(1,1)+\zeta_k$. $\zeta_k$ being a random variable of small amplitude generated to tune the value of MOU. Target matrices are computed as in (\ref{eq1}) for EJD case. The number of matrices is set to $K=5$. For AJD case, these matrices are corrupted by additive noise as given in (\ref{eq2}). The perturbation level is measured by the ratio between the norm of exact term and the norm of disturbance term (i.e. a dual of signal to noise ratio) expressed in dB as: 
\begin{equation}
\mbox{PL(dB)}=10 \log_{10}\left(\frac{\left\|\textbf{A} \textbf{D}_k \textbf{A}^H\right\|_F}{\left\|\Xi_k\right\|_F}\right)
\label{levelp}
\end{equation} 
The error matrix $\Xi_k$ is generated as:
\begin{equation}
\Xi_k=\beta_k \textbf{N}_k
\label{perturbation}
\end{equation}
where $\textbf{N}_k$ is a random perturbation matrix (regenerated at each Monte Carlo run) and $\beta_k$ is a positive number allowing us to tune the perturbation level.\\ The simulation results (i.e performance index) are averaged over $200$ Monte Carlo runs for small dimensional matrices and $20$ Monte Carlo runs for large dimensional matrices.

%%-----------------------------------------------------
\subsection{Exact joint diagonalizable matrices }
%%-----------------------------------------------------
The exact joint diagonalizable matrices are generated as given in equation (\ref{eq1}). This part illustrates the convergence rate of each algorithm where two scenarios are considered. The first one is for small matrix dimension and the second one treats large dimensional matrices. Obtained results are given in the two following points.   
%\begin{enumerate}
\subsubsection{Small matrix dimension}

\begin{figure}
	  \begin{center}
	  \includegraphics[scale=0.5]{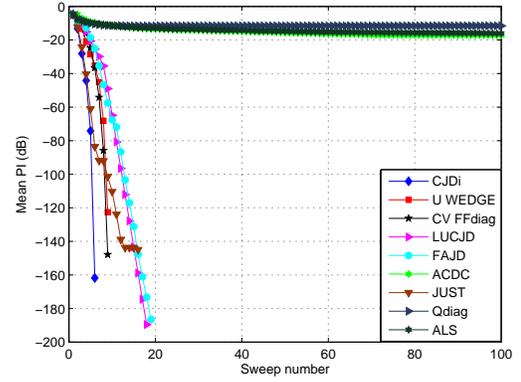}         
    \caption{Mean performance index versus sweep number (matrix dimension $5$, MOU$\mbox{ }< 0.6$, $\mbox{cond}(\textbf{A})<5$).}
	  \label{fig1}
		\end{center}
\end{figure}

\begin{figure}
	  \begin{center}
	  \includegraphics[scale=0.5]{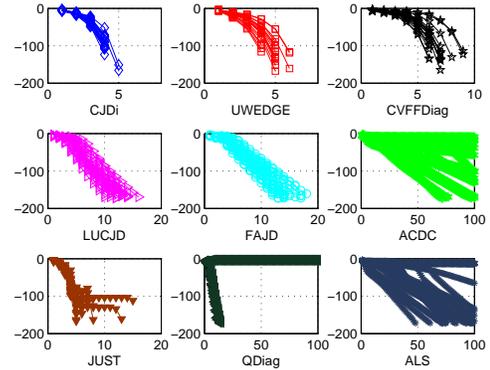}         
    \caption{Different PI of twenty realizations versus sweep number (matrix dimension $5$, MOU$\mbox{ }< 0.6$, $\mbox{cond}(\textbf{A})<5$).}
	  \label{fig1b}
    \end{center}
\end{figure}

\begin{figure}
	  \begin{center}
	  \includegraphics[scale=0.5]{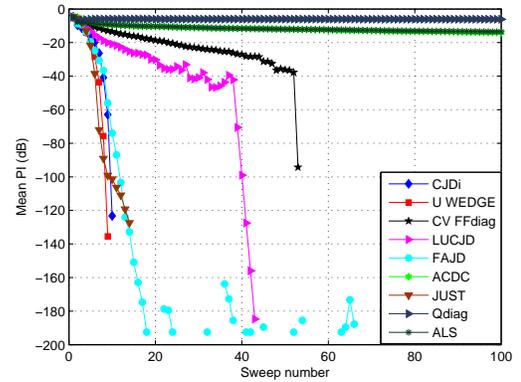}         
    \caption{Mean performance index versus sweep number (matrix dimension $5$, MOU$\mbox{ }< 0.6$, $\mbox{cond}(\textbf{A})>100$).}
	  \label{fig2}
	  \end{center}
\end{figure}

\begin{figure}
	  \begin{center}
	  \includegraphics[scale=0.5]{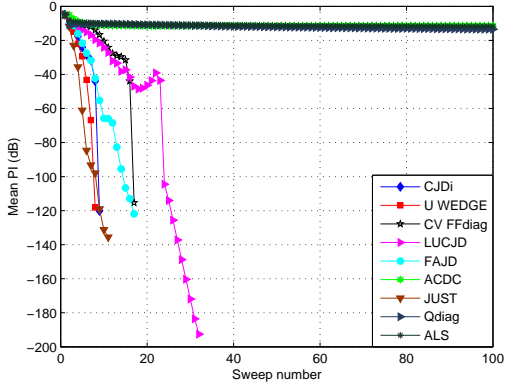}         
    \caption{Mean performance index versus sweep number (matrix dimension $5$, MOU$\mbox{ }< 0.6$, $\mbox{cond}(\textbf{D}_k)>10^4$).}
	  \label{fig3}
	  \end{center}
\end{figure}

\begin{figure}
	  \begin{center}
	  \includegraphics[scale=0.5]{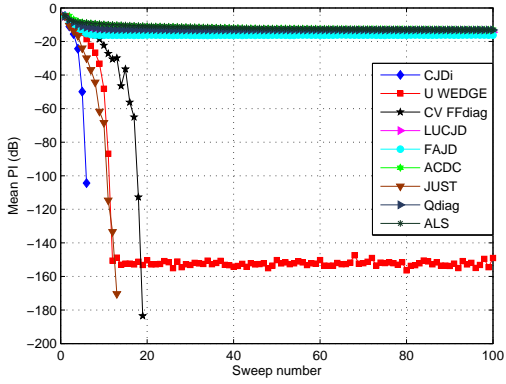}         
    \caption{Mean performance index versus sweep number (matrix dimension $5$, MOU$\mbox{ }> 1-10^{-6}$, $\mbox{cond}(\textbf{A})<5$).}
	  \label{fig4}
	  \end{center}
\end{figure}

For small matrix dimension, four simulations are realized according to the experiments scheme shown in figure \ref{schema} and results are given in figures \ref{fig1}, \ref{fig2}, \ref{fig3} and \ref{fig4}. The first one is the reference case where MOU$\mbox{ }< 0.6$ and Cond($\textbf{A}$)$\mbox{ }<5$. In this case, the majority of algorithms converge at different rates. The fastest one is our developed algorithm where it needs less than ten sweeps to converge. It is followed by CVFFDiag, UWEDGE, JUST, LUCJD and FAJD. 
Note that ACDC, ALS and QDiag diverge in some realizations or need more than $100$ sweeps to converge as illustrated in figure \ref{fig1b} where twenty runs are plotted for each algorithm. % these algorithms are sensitive to the problems matrices.  
%%--reference case ---------------------------------------------------
%%-----------------------------------------------------

%%--reference case ---------------------------------------------------
%%-----------------------------------------------------

Figure \ref{fig2} represents the simulation results in the case of Cond($\textbf{A}$)$\mbox{ }>100$. We observe that UWEDGE, JUST, CJDi and FAJD keep approximatively the same behaviour as before while CVFFDiag and LUCJD became slower in terms of convergence rate. 
%%---ill conditioned mixing matrix-------------------------------------
%%-----------------------------------------------------

In the third simulation where Cond$(\textbf{D}_k)>10^4$, the same remarks as in the previous one can be observed in figure \ref{fig3}. In this simulation UWEDGE and JUST are slightly faster than our proposed algorithm CJDi. 
%%----Ill conditionned Dk-------------------------------------------------
%%-----------------------------------------------------

In the fourth simulation, all parameters are  generated as in the reference case except for the diagonal matrices which are generated  by keeping the MOU greater than $1-10^{-6}$. As shown in figure \ref{fig4}, our proposed algorithm CJDi gives the best results in terms of convergence rate and JD quality and JUST, UWEDGE and CVFFDiag performance is slightly decreased. Otherwise, LUCJD and FAJD diverge in some realization or need more than $100$ sweep to converge hence their performance is degraded considerably.

%%----MOU close to the one--------------------------------------------
%%-----------------------------------------------------

\subsubsection{Large matrix dimension} 

\begin{figure}
	  \begin{center}
	  \includegraphics[scale=0.5]{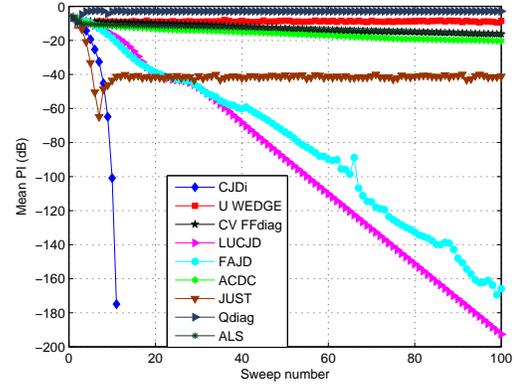}         
    \caption{Performance index versus sweep number (matrix dimension $N= 50$, MOU$\mbox{ }\approx 0.9$, $\mbox{cond}(\textbf{A})<50$).}
	  \label{fig5}
	  \end{center}
\end{figure}

\begin{figure}
	  \begin{center}
	  \includegraphics[scale=0.5]{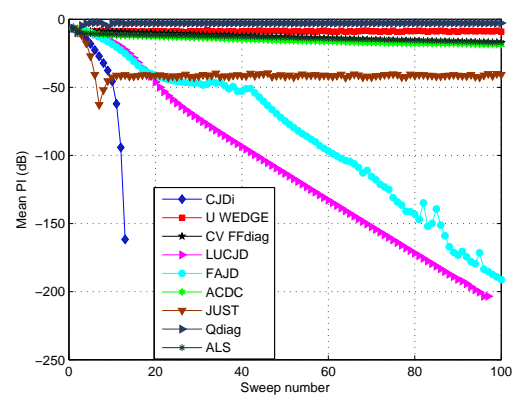}         
    \caption{Mean performance index versus sweep number (matrix dimension $50$, MOU$\mbox{ }\approx 0.9$, $\mbox{cond}(\textbf{A})>100$).}
 		\label{fig6}
 		\end{center}
\end{figure}

\begin{figure}
	  \begin{center}
	  \includegraphics[scale=0.5]{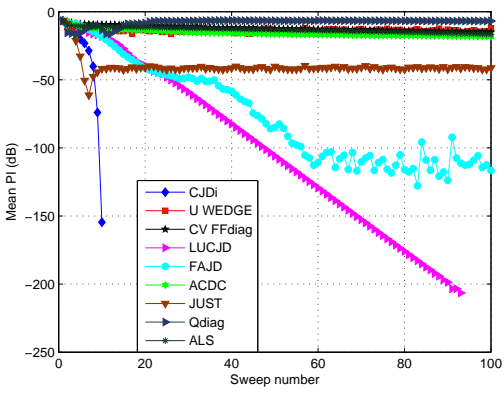}         
    \caption{Mean performance index versus sweep number (matrix dimension $50$, MOU$\mbox{ }\approx 0.9$, $\mbox{cond}(\textbf{A})<50$ , $\mbox{cond}(\textbf{D}_k)>10^4$).}
 		\label{fig6b}
 		\end{center}
\end{figure}

\begin{figure}
	  \begin{center}
	  \includegraphics[scale=0.5]{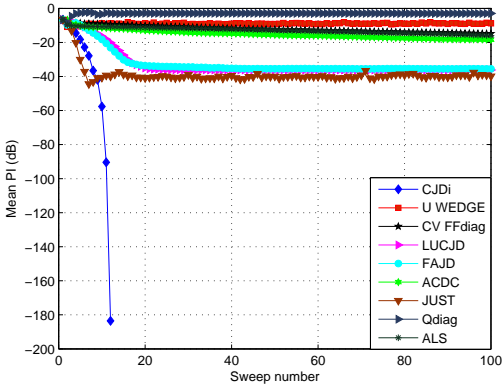}         
    \caption{Mean performance index versus sweep number (matrix dimension $50$, MOU$\mbox{ }\approx 1-10^{-6}$, $\mbox{cond}(\textbf{A})<50$).}
 		\label{fig6c}
 		\end{center}
\end{figure}

In this second experiments set, we have kept fixed the number of matrices as $5$ and matrix dimension is increased to $50$. Four simulation cases are considered as in the small matrix dimension context and the results are given in figures \ref{fig5}, \ref{fig6}, \ref{fig6b} and \ref{fig6c}. 	 
%%----------------reference case -------------------------------------
%%-----------------------------------------------------

The first simulation corresponds to the reference case where Cond($\textbf{A}$)$\mbox{ }<50$, MOU$\mbox{ }\approx 0.9$ and Cond($\textbf{D}_k$)$\mbox{ }<50$. Note that, in addition to ALS, QDiag and ACDC,  UWEDGE and CVFFDiag diverge in some realization or need more than 100 sweep to converge. Hence their averaged performance degraded considerably  in terms of convergence rate and JD quality. JUST has preserved its convergence rate but it has lost its JD quality. LUCJD and FAJD have convergence rates decreased while our algorithm CJDi has preserved its convergence rate and JD quality. 

The same remarks can be observed in the second and third simulations from figures \ref{fig6} and \ref{fig6b} respectively.

In the fourth simulation, the results given in figure \ref{fig6c} show that LUCJD and FAJD have lost their JD quality. Note that only our algorithm CJDi keeps its high performance in terms of convergence rate and JD quality.  
%%----------------Ill conditioned A ----------------------------------
%%-----------------------------------------------------

%%---------------Ill conditioned Dk--------------------------------------
%%-----------------------------------------------------
%%----------------MOU closer to one ---------------------------------
%%-----------------------------------------------------

As a conclusion for EJD case, one can say that our proposed algorithm leads to the best results in major of the  simulation cases and presents a good robustness in the considered adverse scenarios. %\\ In small matrix dimension, UWEDGE presents a slight better performance when mixing or diagonal matrices are ill conditioned however the latter algorithm loses its performance in large matrix dimension.\\ For large matrix dimension, only our algorithm CJDi preserves its performances when the others decreases in convergence rate or in JD quality. 

%%-----------------------------------------------------
\subsection{Approximate joint diagonalizable matrices}
%%-----------------------------------------------------
In this part, we investigates the algorithms robustness to the noise effect.  Simulation scenarios are the same as in exact joint diagonalizable matrices.% Two points are considered where the first one treats small matrix dimension and the second one is reserved to large matrix dimension. Each point contains four simulation scenarios, the first scenario is the reference case where all parameters are generated in good conditions, the second scenario is about ill conditioned mixing matrix, the third scenario generates ill conditioned diagonal matrices and the fourth scenario generates diagonal matrices by keeping the MOU closer to one value.   
 
%\begin{enumerate}
\subsubsection{Small matrix dimension} 
%%---------------reference case --------------------------------------
\begin{figure}
	  \begin{center}
	  \includegraphics[scale=0.5]{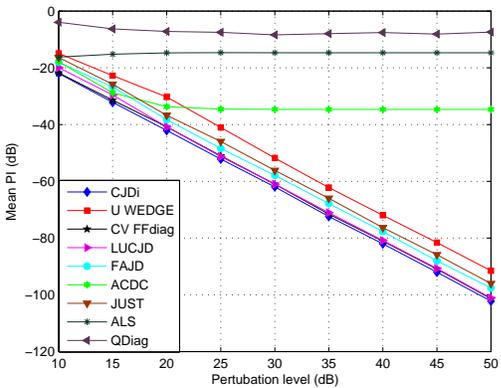}         
    \caption{Mean performance index versus perturbation levels ( matrix dimension $5$, MOU$\mbox{ }< 0.6$, $\mbox{cond}(\textbf{A})<5$)}
	  \label{fig7}
	  \end{center}
\end{figure}
%%-----------------------------------------------------

Obtained results from the first simulations are illustrated in figure \ref{fig7} where plots of mean PI, obtained after $100$ sweep, versus perturbation level are represented for each algorithm. In this scenario, the majority of algorithms converge with different JD quality. Note that, similarly to the EJD case, QDiag, ALS and ACDC diverge in some realization or need more than $100$ to converge.% hence these algorithms are more sensitive to the problem's matrices.
 In that experiment, the CJDi has the best results in terms of JD quality followed by CVFFDiag, LUCJD, FAJD, JUST and UWEDGE. 
%%--------------Ill conditioned A---------------------------------------
\begin{figure}
	  \begin{center}
	  \includegraphics[scale=0.5]{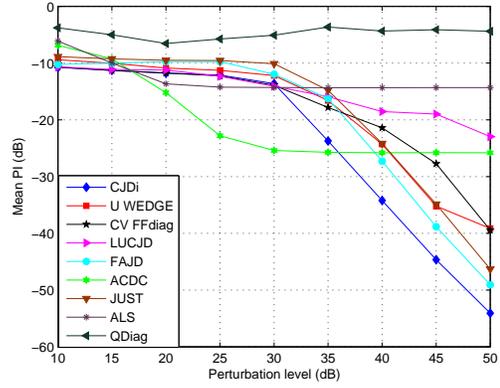}         
    \caption{Mean performance index versus perturbation  levels (matrix dimension $5$, MOU$\mbox{ }< 0.6$, $\mbox{cond}(\textbf{A})>100$).}
	  \label{fig8}
	  \end{center}
\end{figure}
%%-----------------------------------------------------

Results from the second scenario (i.e. Cond($\textbf{A}$)$\mbox{ }>100$) are illustrated in figure \ref{fig8}.  Note that algorithms performance are degraded as compared to the reference case given in figure \ref{fig7}. ACDC and ALS provide the best results when the noise power is high. However, when the latter decreases our proposed algorithm provides the best results.   
%%---------------Ill conditioned Dk--------------------------------------
\begin{figure}
	  \begin{center}
	  \includegraphics[scale=0.5]{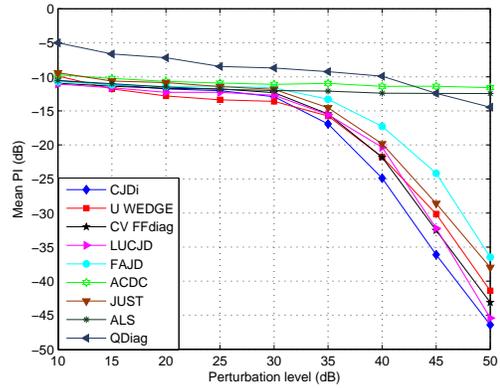}         
    \caption{Mean performance index versus perturbation level (matrix dimension $5$, MOU$\mbox{ } < 0.6$, $\mbox{cond}(\textbf{D}_k)>10^4$).}
	  \label{fig9}
	  \end{center}
\end{figure}
%%-----------------------------------------------------

In the third simulation scenario (Cond($\textbf{D}_k$)$\mbox{ }>10^4$), obtained results are presented in figure \ref{fig9}. Note that our algorithm gives the best results especially for low noise power levels. 
%%---------------MOU closer to the unity--------------------------------------
\begin{figure}
	  \begin{center}
	  \includegraphics[scale=0.5]{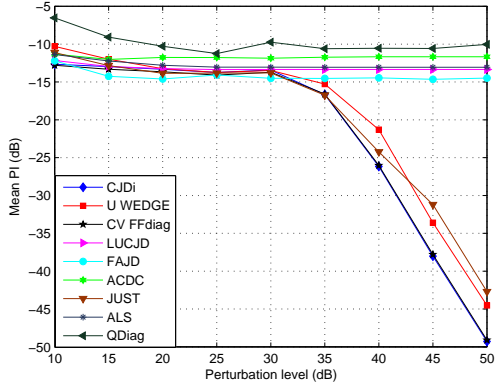}         
    \caption{Mean performance index versus different perturbation levels (matrix dimension $5$, MOU$\mbox{ }> 1-10^{-6}$, $\mbox{cond}(\textbf{A})<5$)).}
	  \label{fig10}
	  \end{center}
\end{figure}
%%-----------------------------------------------------

In the last simulation scenario (MOU$\mbox{ }>1-10^{-6}$), figure \ref{fig10} show that our algorithm and CVFFDiag lead to the best performance in terms of JD quality.  

%-----------------------------------------------------------------------------
\subsubsection{Large matrix dimension}
\begin{figure}
	  \begin{center}
	  \includegraphics[scale=0.5]{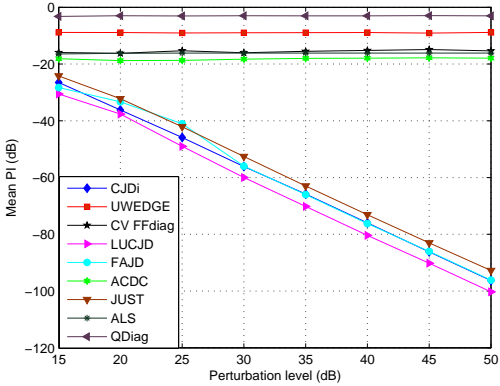}         
    \caption{Mean performance index versus perturbation levels ( matrix dimension $50$, MOU$\mbox{ }\approx 0.9$, $\mbox{cond}(\textbf{A})<50$)}
	  \label{fig11}
	  \end{center}
\end{figure}

\begin{figure}
	  \begin{center}
	  \includegraphics[scale=0.5]{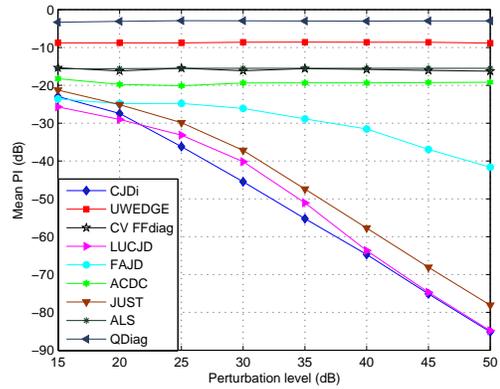}         
    \caption{Mean performance index versus perturbation  levels (matrix dimension $50$, MOU$\mbox{ }\approx 0.9$, $\mbox{cond}(\textbf{A})>100$).}
	  \label{fig12}
	  \end{center}
\end{figure}

\begin{figure}
	  \begin{center}
	  \includegraphics[scale=0.5]{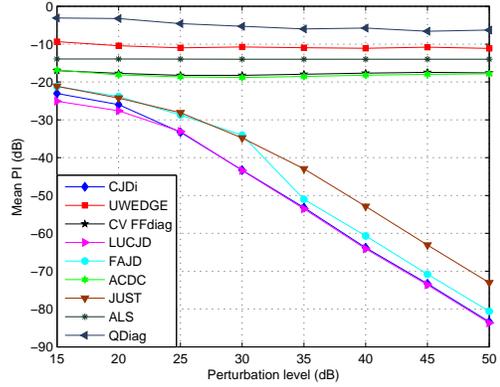}         
    \caption{Mean performance index versus perturbation level (matrix dimension $50$, MOU$\mbox{ }\approx 0.9$, $\mbox{cond}(\textbf{D}_k)>10^4$).}
	  \label{fig13}
	  \end{center}
\end{figure}

\begin{figure}
	  \begin{center}
	  \includegraphics[scale=0.5]{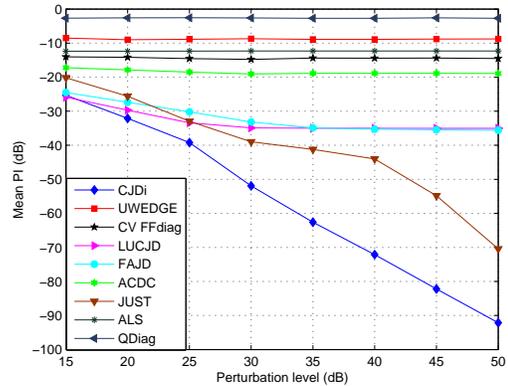}         
    \caption{Mean performance index versus different perturbation levels (matrix dimension $50$, MOU$\mbox{ }> 1-10^{-6}$, $\mbox{cond}(\textbf{A})<50$)).}
	  \label{fig14}
	  \end{center}
\end{figure}

%-----------------------------------------------------------------------------
Here, we consider the JD of $K=5$, $50\times 50$ matrices corrupted by additive noise as given in equation (\ref{eq2}). The four previously mentioned scenarios are considered and the results are given in figures \ref{fig11}, \ref{fig12}, \ref{fig13} and \ref{fig14}. 
%%-----------------reference------------------------------------
%%-----------------------------------------------------

Results from the reference case are given in figure \ref{fig11}. Only LUCJD, CJDi, JUST and FAJD have kept their performance while the others need more than $100$ sweep to converge or diverge in some realizations. Note that LUCJD provides the best results, in that case, followed by CJDi, FAJD and JUST. %This can be explained by  
%%--------------Ill conditioned A---------------------------------------
%%-----------------------------------------------------

results for the second (resp. third) simulation scenarios (Cond($\textbf{A}$)$\mbox{ }>100$ (resp. Cond($\textbf{D}_k$)$\mbox{ }>10^4$)) are given in figure \ref{fig12} (resp. in figure \ref{fig13}). It can be seen that LUCJD still provides the best results followed by CJDi, JUST and FAJD (resp. CJDi, FAJD and JUST) respectively . 
%%---------------Ill condition Dk--------------------------------------
%%-----------------------------------------------------

%In the third simulation scenario, same remarks can be observed in figure \ref{fig13} as the previous cases however LUCJD and CJDi give approximatively the same performance followed by FAJD and JUST.  
%%----------------MOU closer to the unity-------------------------------------
%%-----------------------------------------------------

In the last scenario, given in figure \ref{fig14}, we observed that LUCJD and FAJD have lost their efficiency and that CJDi leads to the best results followed by JUST in that context where MOU$\mbox{ }>1-10^{-6}$. 

All the observed results are summarized in Table \ref{tab_algo} where we express the algorithms sensitiveness to the different factors considered in our study, namely MOU, Cond($\textbf{A}$), Cond($\textbf{D}_k$), $N$ (matrix size) and noise effects. H, M and L refers to high sensitiveness (i.e. high risk of performance loss), moderate sensitiveness and low sensitiveness respectively.

%For approximative joint diagonalizable matrices, simulation scenarios are the same as in the EJD matrices. In this case, plots of mean PI versus perturbation levels are represented in order to illustrate JD algorithm qualities and robustness to the different cited factors. In small matrix dimension our proposed algorithm gives the best results however in large matrix dimension, LUCJD is, in some cases, slightly better than the proposed CJDi algorithm. Once the MOU became closer to one, LUCJD performance is degraded and our algorithm presents the best robustness with respect to adverse MOU values.\\ 
For better results illustration, Table \ref{tab_algo} summarizes all remarks given below.
\begin{table}
\begin{tabular}{|l|c|c|c|c|c|c|}
\hline
 \backslashbox{Alg.}{Fact.} & MOU  & Cond($\textbf{A}$) & Cond($\textbf{D}_k$) &  N & Noise \\% & LUCJD & CJDi
\hline
ACDC & H & L & H & H  & L \\
\hline
ALS & H & L & H & H  & L \\
\hline
QDiag & H & H & H & H & M  \\
\hline
FAJD & H & M & M & L & M  \\
\hline
UWEDGE & L & H & M & L & H \\
\hline
JUST & L & M & M & L & M  \\
\hline
CVFFdiag & H & M & L & H & M  \\
\hline
LUCJD & H & M & L & L & M  \\
\hline
CJDi & L & L & L & L & M  \\
\hline
\end{tabular} %}
\caption{Algorithms performance and sensitiveness to different factors (H, M and L mean High sensitiveness, Moderate sensitiveness and Low sensitiveness, respectively).}
\label{tab_algo}
\end{table}

\section{Conclusion} \label{conclusion}
%------------------------------------------------------------
%------------------------------------------------------------	
In this paper, a new  NOJD algorithm is developed. First, we have considered a basic generalization of JDi from the real to the complex case. Then, we have proposed  our new new algorithm  referred to as CJDi that takes into account the special structure of the transformed matrices and consequently improves the performance in terms of convergence rate and JD quality.
Finally, the comparative study provided in this paper, illustrates the algorithm's efficiency and robustness in adverse JD conditions based on simulation experiments for both EJD and AJD cases. This performance comparison study reveals many interesting features summarized in Tables \ref{algo_com} and \ref{tab_algo}. Based on this study, one can conclude that the CJDi algorithm has the best performance for a relatively moderate computational cost. On the other hand, ACDC, ALS and QDiag  are the most sensitive ones, diverging in most considered realizations. However, when ACDC or ALS converge, they allow to reach the best (lowest performance index) JD quality values in the noisy case.

%----------------------------
%----------------------------
\begin{appendices}
%----------------------------
%----------------------------
\section{Appendix} \label{appendix_A}
%----------------------------
%----------------------------

Let us consider a real symmetric matrix $\mathcal{M}_k$ given in (\ref{fonc}) to which the two rotation matrices $\textbf{H}^i_j(\theta,y)$ and  $\textbf{H}^{i+N}_{j+N}(\theta,y)$ are applied as in (\ref{trans_1}) leading to matrix $\mathcal{M}''_k$.

The objective is to show that $\mathcal{M}''_k$ preserves the same structure of $\mathcal{M}_k$ given in (\ref{fonc}). Using equation (\ref{trans_1}), only $i^{th}$, $j^{th}$, $(i+N)^{th}$ and $(j+N)^{th}$ rows and columns of $\mathcal{M}_k$ are affected according to\footnote{For convenience, we use MATLAB notations. Also, the proof is given only for the $i^{th}$ row as it is obtained in a similar way for the $j^{th}$ row and by the matrix symmetry for the $i^{th}$ and  $j^{th}$ columns.}:

\begin{equation}
\begin{array}{ll}
\mathcal{M}''_k(i,:)   & = \alpha_{11} \mathcal{M}_k(i,:)   +   \alpha_{12} \mathcal{M}_k(j,:)   \\
\mathcal{M}''_k(i+N,:) & = \alpha_{11} \mathcal{M}_k(i+N,:) +   \alpha_{12} \mathcal{M}_k(j+N,:) \\
\end{array}
\label{app_3}
\end{equation}
where $\alpha_{11}$ and $\alpha_{12}$ are given according to Shear parameter $y$ and Givens rotation angle $\theta$ in (\ref{alpha}).
\begin{equation}
\begin{array}{ccc}
\alpha_{11} & = & \cosh(y) \cos(\theta)-\sinh(y) \sin(\theta) \\
\alpha_{12} & = & \cosh(y) \sin(\theta)+\sinh(y) \cos(\theta)\\
\alpha_{21} & = & \sinh(y) \cos(\theta)-\cosh(y) \sin(\theta)\\
\alpha_{22} & = & \cosh(y) \cos(\theta)+\sinh(y) \sin(\theta) 
\end{array}
\label{alpha}
\end{equation}

Since for all $i$: 
\begin{eqnarray}
\mathcal{M}_k(i,1:N) &=& \mathcal{M}_k(i+N,N+1:2N) \nonumber \\
\mathcal{M}_k(i,N+1:2N) &=& -\mathcal{M}_k(i+N,1:N) \label{structure}
\end{eqnarray}
it comes from equations (\ref{app_3}) and (\ref{structure}) that the same  relations apply for matrix $\mathcal{M}''_k$, i.e.
\begin{eqnarray*}
\mathcal{M}''_k(i,1:N) &=& \mathcal{M}''_k(i+N,N+1:2N) \\
\mathcal{M}''_k(i,N+1:2N) &=& -\mathcal{M}''_k(i+N,1:N) 
\end{eqnarray*}

In the second part of the  lemma, we consider  the rotation matrices $\textbf{H}^i_{j+N}(\theta,y)$ and  $\textbf{H}^{j}_{i+N}(\theta,-y)$ and the transformation given in (\ref{trans_2}).

In that case, the $i^{th}$ and $(i+N)^{th}$ rows of $\mathcal{M}'''_k$ can be expressed as:
\begin{equation}
\begin{array}{ll}
\mathcal{M}'''_k(i,:)   & = \alpha_{11} \mathcal{M}_k(i,:)   +   \alpha_{12} \mathcal{M}_k(j+N,:)   \\
\mathcal{M}'''_k(i+N,:) & = -\alpha_{12} \mathcal{M}_k(j,:)   +   \alpha_{11} \mathcal{M}_k(i+N,:) \\
\end{array}
\label{app2_3}
\end{equation}

Again, by taking into account the structure in (\ref{structure}) and equation (\ref{app2_3}), one can obviously observe that $\mathcal{M}'''_k$ has the same structure as $\mathcal{M}_k$.

%%----------------------------
\section{Appendix} \label{appendix_B} 
%%----------------------------
Consider transformations (\ref{trans_1}) and (\ref{trans_2}) where we get $\mathcal{M}''_k$ and $\mathcal{M}'''_k$, respectively. Thanks to Lemma \ref{lemme_1}, these obtained matrices have the structure given in (\ref{fonc}). \\
To compute the simplified JD criterion $\mathcal{C}'_{ij}$ for (\ref{trans_1}) and (\ref{trans_2}), we consider only, the entries twice affected by the considered rotations which are $(i,j)^{th}$, $(i+N,j+N)^{th}$, $(i,j+N)^{th}$, $(j,i+N)^{th}$, $(i,i+N)^{th}$ and $(j,j+N)^{th}$ entries. \\
Considering the structure given in (\ref{fonc}): 
\begin{itemize}
	\item $(i,i+N)^{th}$ and $(j,j+N)^{th}$ entries are equal to zero;
	\item $(i+N,j+N)^{th}$ entries are equal to $(i,j)^{th}$ entries.
\end{itemize}
The development of  considered transformations, $(i,j+N)^{th}$ and $(j,i+N)^{th}$ entries are not changed and we get :
$$ \mathcal{M}''_k(i,j+N) = \mathcal{M}_k(i,j+N)$$
$$ \mathcal{M}'''_k(i,j) = \mathcal{M}_k(i,j)$$
Combining these results with equation (\ref{ciNjN}) (resp. (\ref{cijN})), the simplified JD criteria $\mathcal{C}'_{ij}$ for matrix transforms (\ref{update}) and (\ref{trans_1}) considering ($i$,$j$) indices (resp. for matrix transforms (\ref{update}) and (\ref{trans_2}) considering ($i$,$j+N$) indices) are equal up to a constant factor.

\end{appendices}

\section*{Acknowledgment}

The authors would like to express  sincere gratitude to Dr. Xiao-Feng Gong, his student Ke Wang and Dr. Redha Iferroudjene for providing us with MATLAB programs of their respective NOJD methods in \cite{feng} and \cite{JUST}.

\addcontentsline{toc}{chapter}{references}
\bibliographystyle{IEEEbib}
\bibliography{mybib2}
%\end{onecolumn}
\end{document}